\documentclass[preprint]{aastex63}

\newcommand{\logg} {\log \textsl{\textrm{g}}}

\newcommand{\Te} {T_{\rm eff} }

\newcommand{\msun} {$M_\odot$}

\newcommand\gta{\lower 0.5ex\hbox{$\buildrel > \over \sim\ $}} 
\newcommand\lta{\lower 0.5ex\hbox{$\buildrel < \over \sim\ $}} 

\usepackage{xcolor}

\usepackage{afterpage}

\submitjournal{AJ}

\shorttitle{ZZ Ceti Stars in the {\it Gaia} Survey}
\shortauthors{Vincent et al.}

\begin{document}

\title{Searching for ZZ Ceti White Dwarfs in the {\it Gaia} Survey}

\author{Olivier Vincent}
\email{ovincent@astro.umontreal.ca}
\affiliation{D\'epartement de Physique, Universit\'e de Montr\'eal,
  C.P.~6128,\\ Succ.~Centre-Ville, Montr\'eal, Qu\'ebec H3C 3J7, Canada}
  
\author{P. Bergeron}
\email{bergeron@astro.umontreal.ca}
\affiliation{D\'epartement de Physique, Universit\'e de Montr\'eal,
  C.P.~6128,\\ Succ.~Centre-Ville, Montr\'eal, Qu\'ebec H3C 3J7, Canada}

\author{David Lafreni\`ere}
\email{david@astro.umontreal.ca}
\affiliation{D\'epartement de Physique, Universit\'e de Montr\'eal,
  C.P.~6128,\\ Succ.~Centre-Ville, Montr\'eal, Qu\'ebec H3C 3J7, Canada}

\begin{abstract}

The {\it Gaia} satellite recently released parallax measurements for
$\sim$260,000 high-confidence white dwarf candidates, allowing for
precise measurements of their physical parameters. By combining these
parallaxes with Pan-STARRS and $u$-band photometry, we measured the
effective temperature and stellar mass for all white dwarfs in the
Northern Hemisphere within 100 parsecs of the Sun, and identified a
sample of ZZ Ceti white dwarf candidates within the so-called
instability strip. We acquired high-speed photometric observations for
90 candidates using the PESTO camera attached to the 1.6-m telescope
at the Mont-M\'egantic Observatory. We report the discovery of 38 new
ZZ Ceti stars, including two very rare ultra-massive pulsators. We
also identified 5 possibly variable stars within the strip, in
addition to 47 objects that do not appear to show any photometric
variability. However, several of those could be variable with an
amplitude below our detection threshold, or could be located outside
the instability strip due to errors in their photometric
parameters. In the light of our results, we explore the trends
of the dominant period and amplitude in the $M - \Te$ plane, and briefly 
discuss the question of the purity of the ZZ Ceti instability strip 
(i.e.~a region devoid of non-variable stars).

\end{abstract}


\section{Introduction}\label{sec:intro}

White dwarf stars represent the end product of 97\% of the stars in
the Galaxy. Their cores no longer produce energy through nuclear
fusion, and so they slowly cool down over the span of billions of
years, allowing us to interpret their temperature sequence as an
evolutionary track. Most white dwarfs go through an instability stage
at some point in their lives, depending on the chemical composition of
their outer stellar envelope, during which they exhibit nonradial
$g$-mode pulsations. For instance, once DA (hydrogen-line) white
dwarfs reach an effective temperature between $\Te\sim 12,300$~K and
$\sim$10,200~K (for a surface gravity of $\logg=8$,
\citealt{gianninas2014}), their internal conditions become prone to
such pulsations, manifesting themselves as periodic variations in the
luminosity of the star, with periods typically ranging from 100~s
\citep{voss2006} to 2000~s \citep{green2015}, and relative amplitudes
from 0.1 to 40\% \citep{mukadam2004}.

One of the main interests surrounding the region in the $\logg - \Te$
plane containing the variable DAs, namely the ZZ Ceti instability
strip, is to determine whether it is pure or not. A pure strip, devoid
of any photometrically constant DA white dwarfs, would suggest that ZZ
Ceti stars represent an evolutionary phase through which most, if not
all, hydrogen-atmosphere white dwarfs are expected to cool. We could
then use asteroseismology as a tool to study the internal structure
not only of ZZ Ceti stars, but also of the population of DA white
dwarfs as a whole \citep{giammichele2017}. On the other hand, an
impure strip containing a mix of variable and non-variable DA stars
would imply a missing parameter in our evolutionary models
\citep{fontaine2008}. The purity of the instability strip has a long
history of swinging back and forth between these two possibilities. On
one hand, there are studies such as that of \citet{gianninas2014}, who
restricted their sample to only the brightest ZZ Ceti stars with high
signal-to-noise spectra, and whose results point toward a pure
instability strip. But there are also many studies claiming the strip
to be populated with both variable and non-variable stars (see, for
example, \citealt{mukadam2005}). In most of those cases, the
photometrically constant stars were either found to be variable when
using better instruments \citep{castanheira2007}, or proven to be 
located outside the strip by measuring their parameters with higher 
quality data \citep{gianninas2005}. Even though it is an uphill battle, 
the consensus seems to be slowly heading toward a pure strip.

Over the years, there have been many efforts to define the
spectroscopic ZZ Ceti instability strip both empirically and
theoretically. The theoretical determination of the strip edges is
still a work in progress, as it strongly depends on the physical
assumptions made in these studies, especially when it comes to the
efficiency of convective energy transport (see \citealt{fontaine2008},
\citealt{althaus2010}, and \citealt{corsico2019} for excellent reviews
on the subject). Furthermore, the assumptions behind the theoretical
edges are often based on their empirical locations, which are
themselves dependent on a variety of factors, such as the
signal-to-noise ratio of the spectra \citep{gianninas2005}. Building a
large, homogeneous sample of photometrically variable and constant
stars inside and near the instability strip is the first step towards
a robust determination of the empirical
edges. \citet{bergeronetal1995} began this venture by collecting
time-averaged optical spectra to measure the $\Te$ and $\logg$ values
of known ZZ Ceti stars, allowing them to select new ZZ Ceti candidates
with high confidence. Since then, this so-called spectroscopic
technique has been used repeatedly to identify new ZZ Ceti stars, with
perhaps the most impressive of these studies being that of
\citet{mukadam2004} who reported in a single paper the discovery of 35
new ZZ Ceti stars in the Sloan Digital Sky Survey (SDSS) and the
Hamburg Quasar Survey. In parallel, the same approach has been used to
constrain the exact location of the ZZ Ceti instability strip by also
studying non-variable DA white dwarfs around the strip \citep[see,
  e.g.][]{gianninas2005}. By far, the spectroscopic technique has been
the go-to method to identify new candidates, being one of the main
contributors of the $\sim$200 new ZZ Ceti stars found in the last 20
years or so \citep{bognar2016}.

Unfortunately, the determination of the exact location of the
empirical ZZ Ceti instability strip has been hampered by the use of
different model atmospheres in these spectroscopic investigations,
which differ in terms of the Stark broadening theory for the hydrogen
lines, as well as different assumptions about the convective
efficiency. More importantly, \citet{tremblay2011} demonstrated that
the mixing-length theory used to describe the convective energy
transport in previous model atmosphere calculations was responsible
for the apparent increase of spectroscopic $\logg$ values below
$\Te\sim13,000$~K, a problem that could be solved by relying on
realistic 3D hydrodynamical model atmospheres
\citep{tremblay2013}. Given this confusing situation, we decided to
revisit this problem more quantitatively in a homogeneous fashion.

\begin{figure}
\begin{center}
 \includegraphics[angle=270,width=0.9\columnwidth,clip,trim=1in 0.51in 0.5in 0]{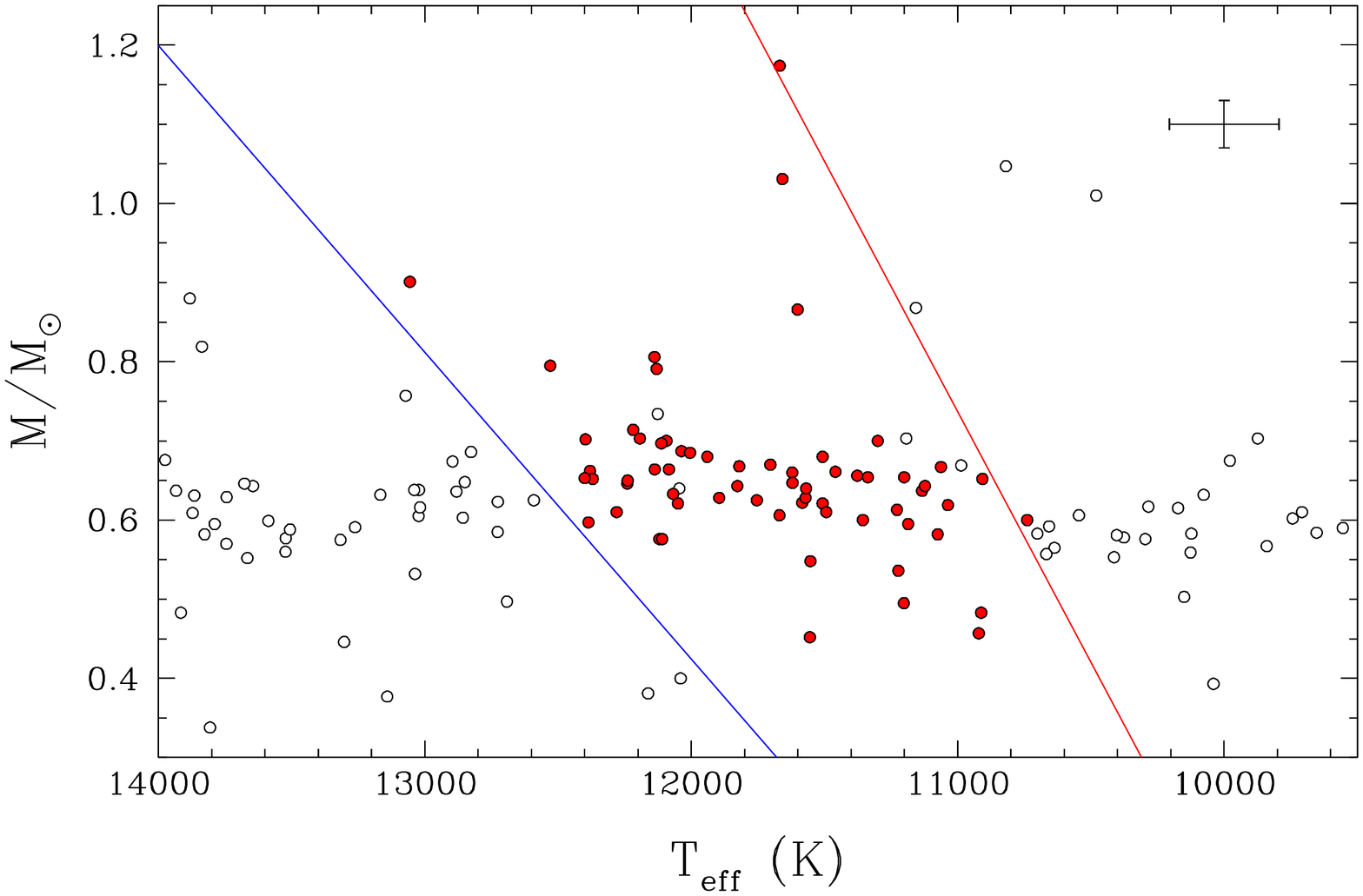}
  \caption{Distribution of the ZZ Ceti stars (red) and photometrically
    constant white dwarfs (white) from \citet{green2015} in the $M -
    \Te$ plane. Here the spectroscopic parameters have been corrected
    for hydrodynamical 3D effects. The cross in the upper right corner
    represents the average uncertainties in both parameters. The
    empirical ZZ Ceti instability strip is indicated by the blue (hot
    edge) and red (cool edge) lines.}
   \label{fig:zzstripgreen}
\end{center}
\end{figure}

Our starting point is the study of \citet{green2015} who presented new
high-speed photometric observations of ZZ Ceti white dwarf candidates
drawn from the spectroscopic survey of bright DA stars in the
Villanova White Dwarf Catalog \citep{mccook1999} by
\citet{Gianninas2011}, and from the spectroscopic survey of white
dwarfs within 40 parsecs of the Sun by \citet{limoges2015}. Figure 2
of Green et al.~summarizes the distribution of $\logg$ as a function
of $\Te$ for all ZZ Ceti and photometrically constant white dwarfs in
their sample, providing us with an empirical instability strip based
on the largest (and mostly) homogeneous sample yet. However, their
spectroscopic solutions, obtained from model atmospheres based on the
ML2/$\alpha=0.7$ version of the mixing-length theory, were not
corrected for hydrodynamical 3D effects. Here we first apply the 3D
corrections from \cite{tremblay2013} to the spectroscopic $\Te$ and
$\logg$ values, and then convert the $\logg$ values into stellar
masses ($M$) using evolutionary models described in Section
\ref{sec:sample}. The resulting distribution of white dwarfs in the $M
- \Te$ plane is displayed in Figure \ref{fig:zzstripgreen}. We use
these results to derive improved empirical boundaries for the ZZ Ceti
instability strip, also reproduced in Figure \ref{fig:zzstripgreen},
which will serve as a reference in our discussion below. The 3D
hydrodynamical corrections can be neglected in the context of
photometric analyses, as discussed by \citet{tremblay2013}, who showed
that 1D or 3D-corrected models yield similar results for DA white
dwarfs in the 7000-14,000~K temperature range (see their Figure 16).

With the second {\it Gaia} data release, trigonometric parallaxes have
become available for an unprecedented number of white dwarf stars,
opening a whole new window of opportunity to identify new ZZ Ceti
stars. Indeed, distances derived from such parallaxes are an essential
ingredient for precise measurements of their physical parameters using
the so-called photometric approach. In this paper, we make use of the
Panoramic Survey Telescope and Rapid Response System (Pan-STARRS)
photometry for the first time in the context of identifying new ZZ
Ceti stars and constraining the empirical edges of the photometric ZZ
Ceti instability strip. By combining {\it Gaia} astrometric data with
this nearly all-sky photometric survey, at least in the northern
hemisphere, we obtain one of the largest photometric samples of ZZ
Ceti candidates yet. This combination of parallax and photometric data
has been thoroughly investigated by \citet{bergeron2019}, who showed
that physical parameters --- namely $\Te$ and $M$ --- derived from
spectroscopy and photometry reveal systematic offsets (see their
Figure 4). We thus expect the empirical ZZ Ceti instability strip
obtained from our photometric analysis to exhibit similar offsets with
respect to spectroscopic determinations.

Our selection of ZZ Ceti candidates is first discussed in Section
\ref{sec:sample}, while the high-speed photometric follow-up program
for our selected ZZ Ceti candidates, as well as the data reduction
procedure, are described in Section \ref{sec:data}. Our results,
including the discovery of 38 (and possibly 43) new ZZ Ceti stars and
the discussion of the empirical photometric instability strip, are
presented in Section \ref{sec:res}. Our conclusions follow in Section
\ref{sec:conc}.
 
\section{Candidate Selection}\label{sec:sample}

\begin{figure}
\begin{center}
 \includegraphics[angle=270,width=0.9\columnwidth,clip,trim=1in 0.51in 0.45in 0]{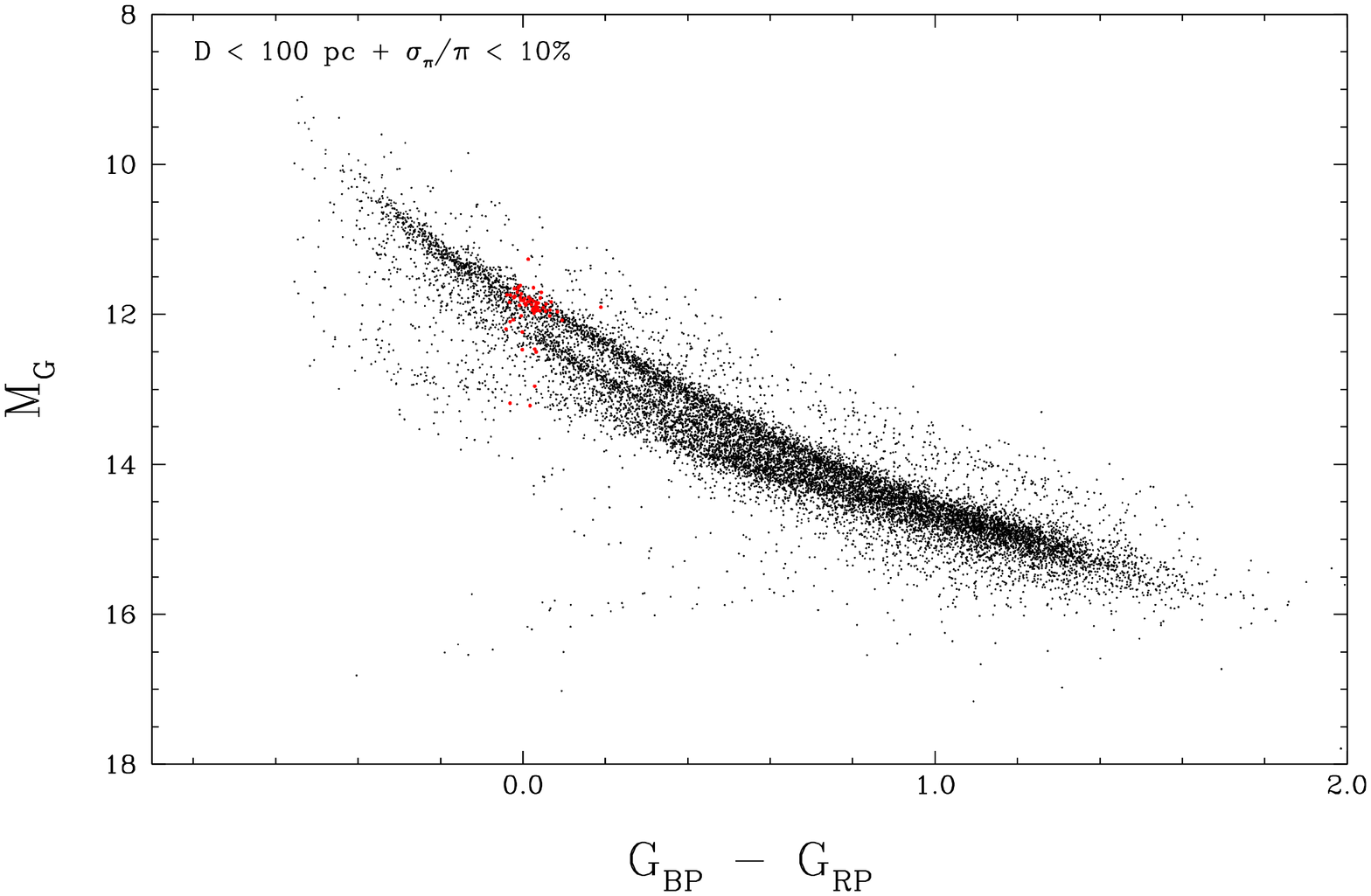} 
 \caption{Color-magnitude diagram for {\it Gaia} white dwarfs and
   white dwarf candidates within 100 pc of the Sun with parallax
   measurements more precise than 10\%. Our search for pulsating ZZ
   Ceti pulsators is based on this parallax- and color-selected sample
   containing 12,857 objects. Previously known ZZ Ceti stars are shown
   in red.}
   \label{fig:hrd}
\end{center}
\end{figure}

Our initial sample consists of all objects from the {\it Gaia} Data
Release 2 \citep{gaia2016,gaia2018} within 100 parsecs of the Sun and
parallax measurements more precise than 10\%. This distance limit was
chosen so that interstellar reddening could be neglected in our
photometric analysis described below \citep{harris2006}. To define our
white dwarf candidate sample, we apply the selection criteria
described in Section 2.1 of \citet{gaia2018hrd} excluding the limits
on \texttt{flux\_over\_error} for $G$, $G_{\rm BP}$, and $G_{\rm BP}$
magnitudes. More specifically, we select objects with an absolute {\it
  Gaia} magnitude $M_G >9$ and color indices $-0.6 \leq G_{\rm BP} -
G_{\rm RP} \leq 2.0$. Figure \ref{fig:hrd} shows the {\it Gaia}
color-magnitude diagram for the 12,857 objects contained in this
initial sample. We note that this selection of white dwarf candidates
excludes the extremely low-mass (ELM) white dwarf pulsators \citep{bell2017}, 
as they are located significantly above the white dwarf sequence in the {\it
  Gaia} color-magnitude diagram \citep{gaia2019}. However, all of the
currently known ELM pulsators have distances of the order of
kiloparsecs \citep{Brown2011}, and their number within 100 pc is
expected to be extremely small \citep{kawka2020}.

We then cross-match this initial sample with the Pan-STARRS Data
Release 1 catalog \citep{chambers2016} using the following
algorithm\footnote{Note that the algorithm described here is being
  used for the Pan-STARRS photometry provided in the Montreal White
  Dwarf Database \citep[MWDD,][]{dufour2017}.}. For each {\it Gaia}
object, a first query is made at the {\it Gaia} coordinates in a 5
arcsecond radius circle, and if only one object is found and has good
quality flags, it is chosen as the cross-match. If no objects are
found, we expand the search query radius to 20 arcseconds. If multiple
Pan-STARRS objects are found within this search radius, the {\it Gaia}
object is looked up on the SIMBAD Astronomical Database
\citep{wenger2000} for SDSS $ugriz$ magnitudes \citep{york2000}. Since
SDSS and Pan-STARRS $griz$ filters are comparable, we use available
SDSS photometry to select the Pan-STARRS object with the closest
matching photometry, allowing a difference of up to 0.3 mag per
filter. In the case where no Pan-STARRS objects meet this criteria,
the cross-match fails. If no SDSS photometry is available, we use
instead the $G-r$ relationship described in \citet{evans2018} to
estimate a SDSS $r$ magnitude and select the object with the closest
Pan-STARRS $r$ magnitude, up to a difference of 0.7 mag.

With the {\it Gaia} parallaxes and Pan-STARRS $grizy$ photometry in
hand, every object in our initial sample is fitted using the
photometric technique described at length in \citet{bergeron1997},
together with the pure hydrogen\footnote{Worth mentioning in the
  present context, the pure hydrogen model atmospheres --- calculated
  with the ML2/$\alpha=0.7$ version of the mixing-length theory ---
  are identical to those used to determine the empirical ZZ Ceti strip
  based on spectroscopy displayed in Figure \ref{fig:zzstripgreen}.}
and pure helium model atmospheres discussed in \citet{bergeron2019}
and references therein. As mentioned above, given the distance limit
of our sample, interstellar reddening is neglected altogether. The
fitted parameters are the effective temperature, $\Te$, and the solid
angle, $\pi (R/D)^2$, where $R$ is the radius of the star and $D$ its
distance from Earth, derived from the trigonometric parallax
measurement. The fitted stellar radii can be converted into surface
gravity ($\logg$) and stellar mass ($M$) using evolutionary
models\footnote{See
  http://www.astro.umontreal.ca/$\sim$bergeron/CoolingModels.} similar
to those described in \citet{fontaine2001} with (50/50) C/O-core
compositions, $q({\rm He})\equiv M_{\rm He}/M_{\star}=10^{-2}$, and
$q({\rm H})=10^{-4}$ or $10^{-10}$ for the pure hydrogen and pure
helium solutions, respectively. As discussed in the Introduction, 3D
hydrodynamical corrections can be neglected in the context of
photometric analyses \citep{tremblay2013}.

\begin{figure}
\begin{center}
 \includegraphics[width=0.9\columnwidth,clip=True,trim=0 3.5in 0 3.2in]{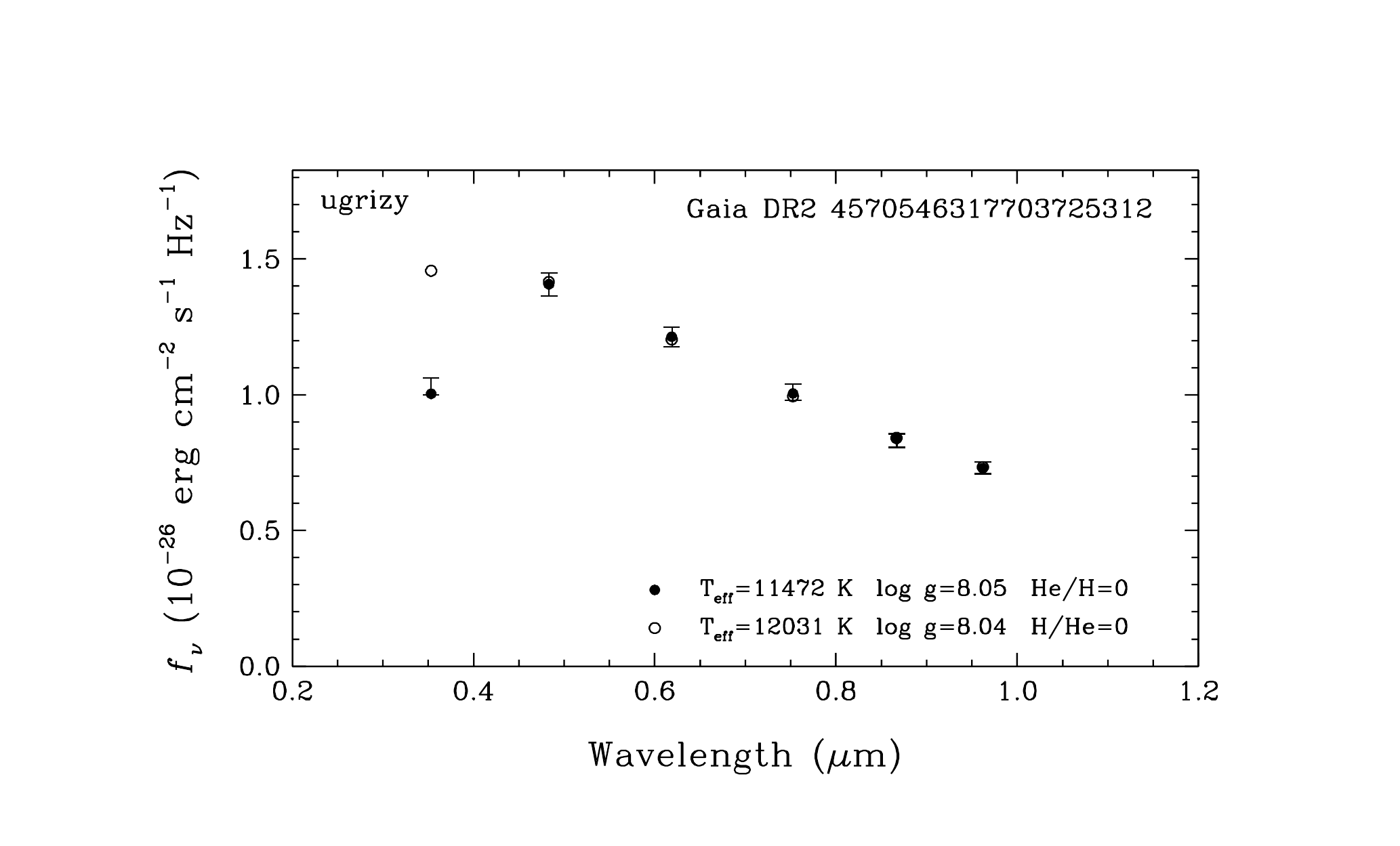}
 \caption{Sample photometric fit to a ZZ Ceti white dwarf candidate
   using Pan-STARRS $grizy$ and CFIS-$u$ photometry (error bars),
   combined with the {\it Gaia} parallax measurement. Filled circles
   correspond to our best fit under the assumption of a pure hydrogen
   atmospheric composition, while the open circles assume a pure
   helium atmosphere. Note that the CFIS-$u$ data point is not used in
   these fits and serves only to discriminate between the pure
   hydrogen and pure helium solutions (see text); the results clearly
   indicate that this object is a hydrogen-atmosphere white dwarf.}
   \label{fig:fitphoto}
\end{center}
\end{figure}

In Figure \ref{fig:fitphoto}, we show a typical fit for one object in
our sample using Pan-STARRS $grizy$ photometry and the {\it Gaia}
parallax measurement. As can be seen from this result, hydrogen- and
helium-atmosphere white dwarfs can be difficult to distinguish based
on Pan-STARRS $grizy$ photometry alone, as their average flux
distribution in the 0.4-1.0 $\mu$m tends to be quite similar. To
overcome this problem, we supplement our set of $grizy$ photometry
with $u$-band photometry, if available, taken from the SDSS or from
the ongoing Canada-France Imaging Survey (CFIS) described in
\citet{ibata2017}. The wavelength coverage of the $u$ bandpass
includes the Balmer jump, which is a very distinctive feature between
hydrogen- and helium-atmosphere white dwarfs. Indeed,
hydrogen-atmosphere white dwarfs have a significant drop in $u$-band
flux, whereas their helium-atmosphere counterparts have a more
continuous flux distribution. The $u$ magnitude is not included in the
fitting procedure but it is used instead in our analysis (see below)
to discriminate between the pure hydrogen and pure helium solutions,
as illustrated in Figure \ref{fig:fitphoto}, where the drop in the
$u$-flux caused by the Balmer jump is accurately reproduced by the
pure hydrogen model.

The photometric fits are also useful to identify non-white dwarf
objects when the measured parameters are unrealistic, in particular
the stellar radius. However, it is also possible to obtain a bad fit
if our photometric cross-match is erroneous, in which case we may miss
white dwarf candidates in our initial sample. These two scenarios
affected less than 1\% of the objects with a Pan-STARRS cross-match.

\begin{figure}[p]
\begin{center}
 \includegraphics[angle=270,width=0.82\columnwidth,clip,trim=0.95in 0.5in 1in 0.66in]{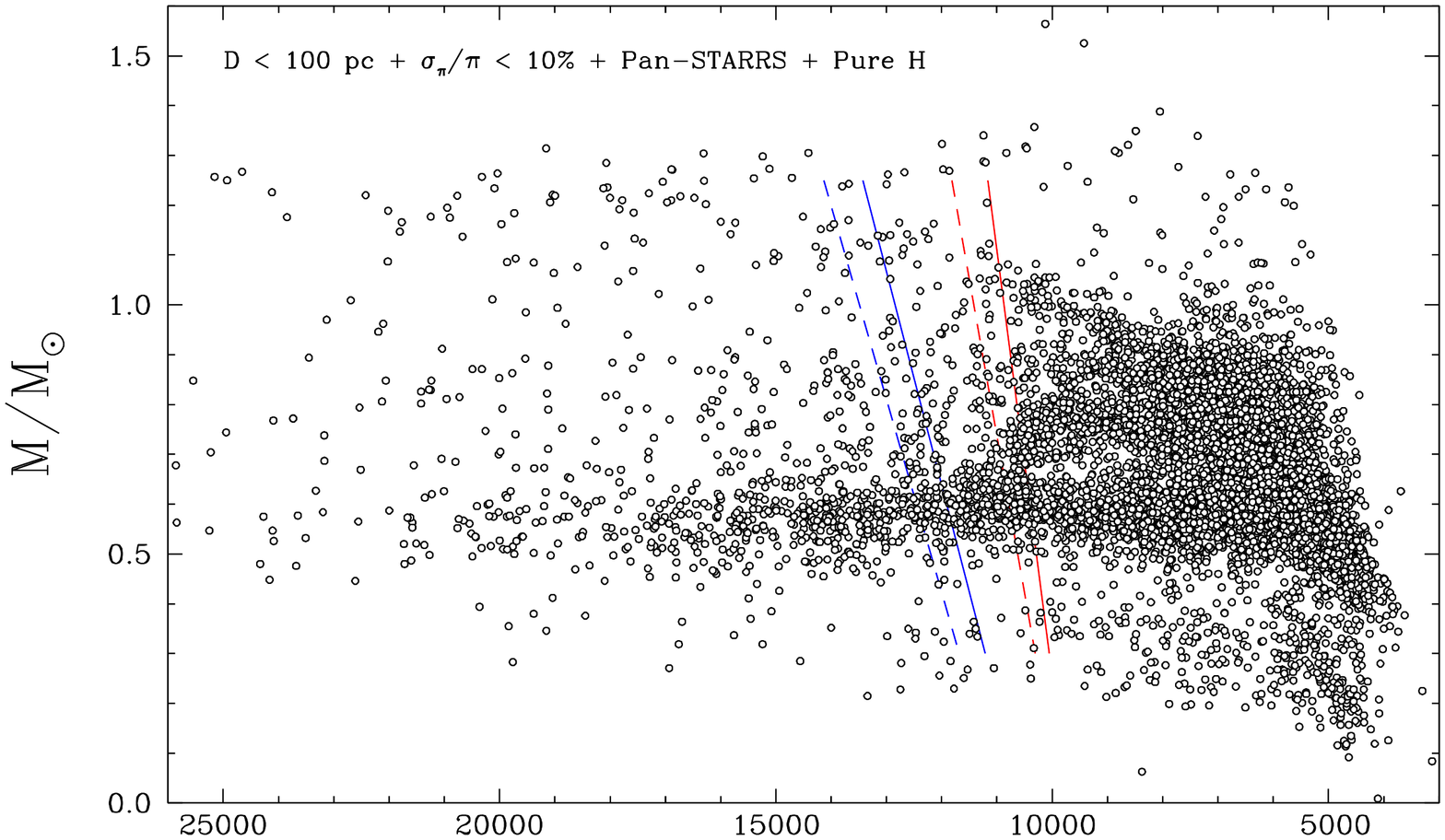}
 \includegraphics[angle=270,width=0.82\columnwidth,clip,trim=1.1in 0.5in 0.45in 0.66in]{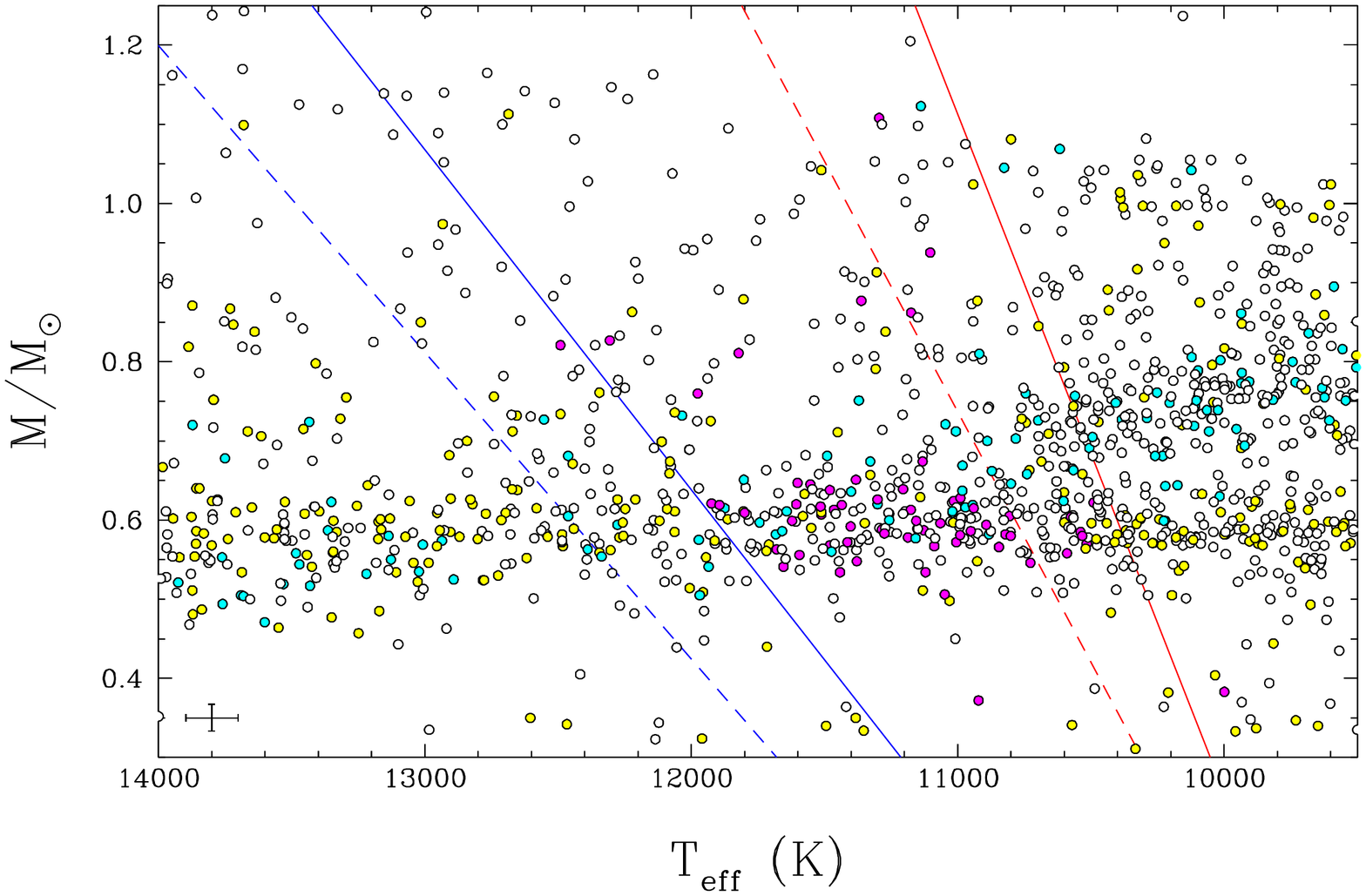}
 \caption{Top: Distribution of the objects in our sample in the $M -
   \Te$ plane, measured using the photometric technique assuming pure
   hydrogen atmospheres. The spectroscopic (dashed lines) and
   photometric (solid lines) empirical ZZ Ceti instability strips are
   indicated by the blue (hot edge) and red (cool edge) lines. Bottom:
   Same as the top panel, but zoomed in on the instability strip; the
   cross in the lower left corner represents the average uncertainties
   in both parameters. Known ZZ Ceti (magenta), DA (yellow) and non-DA
   (cyan) white dwarfs are also identified.}
   \label{fig:fullsamp}
\end{center}
\end{figure}

The stellar masses for all objects in our sample are displayed in
Figure \ref{fig:fullsamp} as a function of effective temperature; here
a pure hydrogen atmospheric composition is assumed for all
objects. The upper panel shows the full $M - \Te$ distribution of our
sample. As discussed in detail by \citet{bergeron2019}, the large
masses observed below $\Te\sim10,000$~K correspond to
helium-atmosphere white dwarfs containing small traces of hydrogen (or
carbon and other heavy elements), whose masses are overestimated when
analyzed with pure hydrogen or even pure helium model atmospheres.

\begin{figure}
\begin{center}
 \includegraphics[width=0.9\columnwidth,clip,trim=0in 3.5in 0in 2.5in]{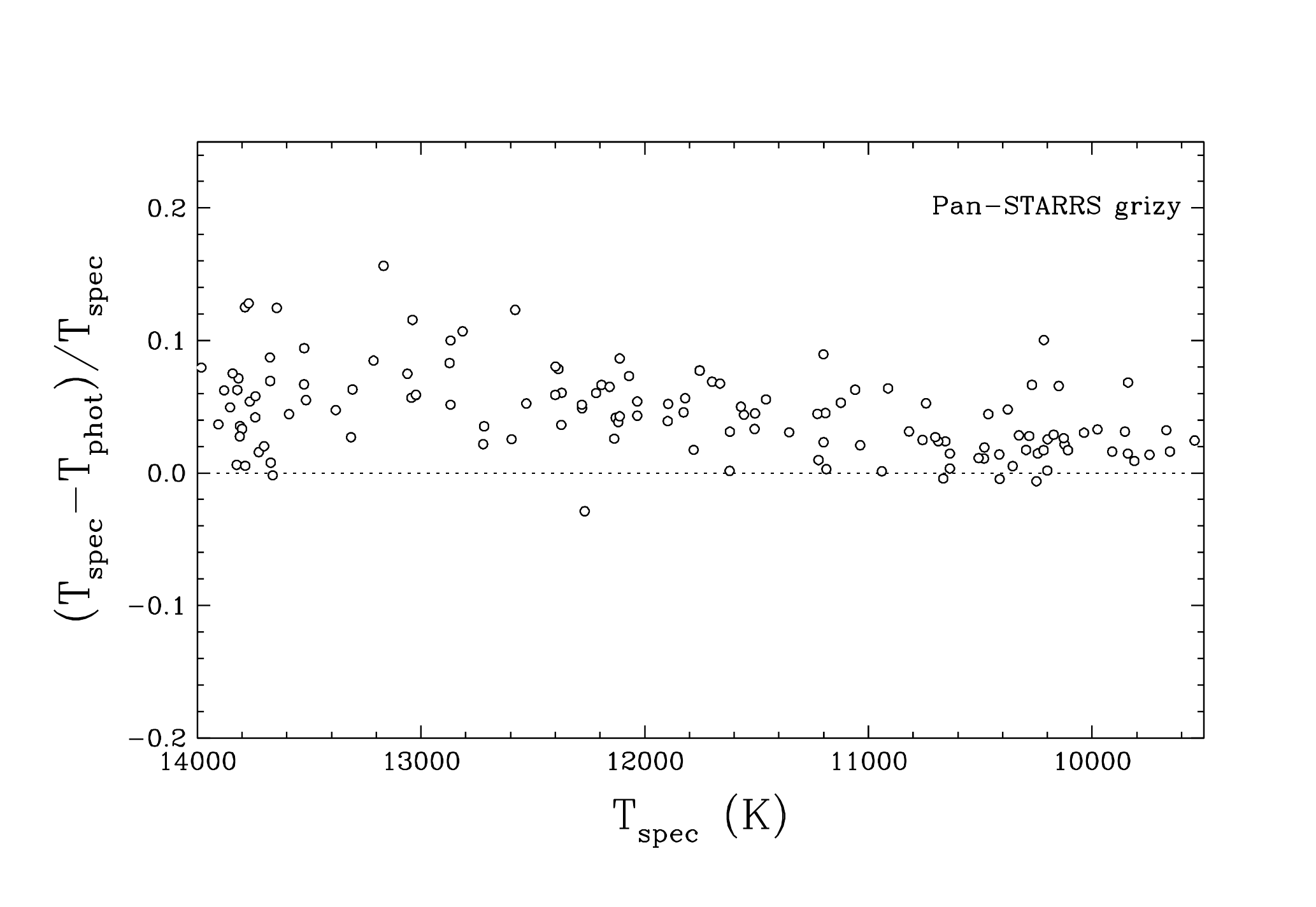}
\caption{Differences between spectroscopic and photometric effective
  temperatures as a function of $\Te$ for DA stars in the region of
  interest, drawn from the sample of \citet{Gianninas2011}, using
  photometric fits to the Pan-STARRS $grizy$ data. The dotted line
  indicates equal temperatures.}\label{fig:comp_VINCENT}
\end{center}
\end{figure}

Of more interest in the present context is the range of effective
temperature where ZZ Ceti white dwarfs are expected, displayed in the
bottom panel of Figure \ref{fig:fullsamp}. Also reproduced in both
panels (dashed lines) is the location of the ZZ Ceti instability strip
determined empirically by \citet[][see Figure
    \ref{fig:zzstripgreen}]{green2015}. In principle, this
instability strip could be used to select our ZZ Ceti candidates for
follow-up high-speed photometry. However, as demonstrated by
\citet{bergeron2019}, photometric temperatures obtained from
Pan-STARRS $grizy$ photometry are significantly lower than
spectroscopic temperatures. We reproduce in Figure
\ref{fig:comp_VINCENT} the results from Bergeron et al.~(their Figure
4) but only for the range of temperature of interest. We can see that
the temperature offset varies slightly as a function of $\Te$, but
that it is otherwise well defined on average. We thus use the results
displayed in Figure \ref{fig:comp_VINCENT} to apply a temperature
correction to the spectroscopic instability strip determined by
  \citet{green2015} to estimate the photometric boundaries of the
strip, as indicated by solid lines in Figure \ref{fig:fullsamp}. This
is the region of the $M - \Te$ plane that will be used to define our
sample of ZZ Ceti candidates.

\begin{figure}
\begin{center}
 \includegraphics[angle=270,width=0.85\columnwidth,clip,trim=1.65in 0.45in 0.55in 0.66in]{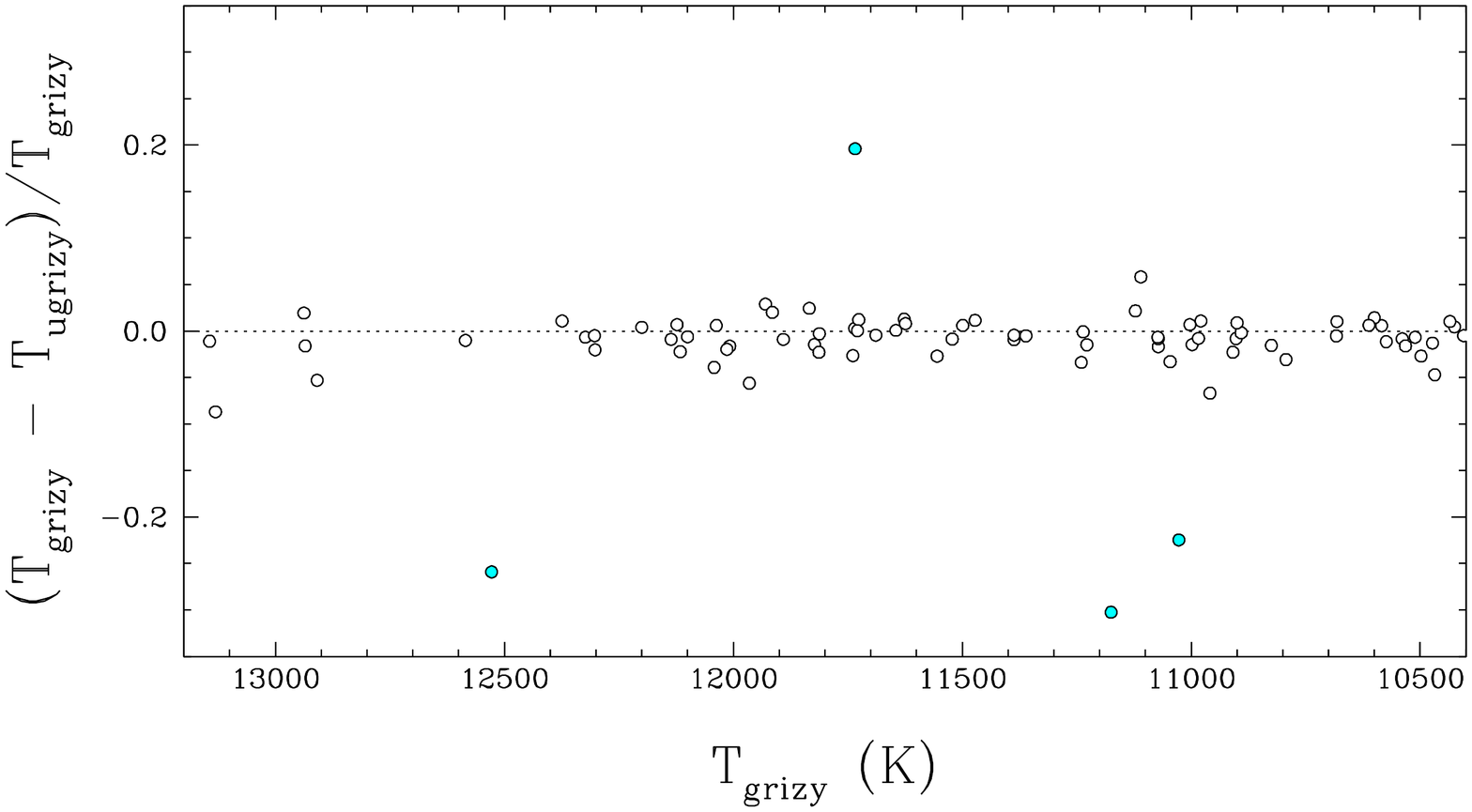}
 \caption{Differences (in \%) between photometric temperatures
   measured using only Pan-STARRS photometry ($T_{\rm grizy}$) and
   those obtained by also including SDSS or CFIS $u$-band photometry
   ($T_{\rm ugrizy}$) for objects within the ZZ Ceti region. The
   dotted line indicates equal temperatures. White and cyan symbols
   correspond, respectively, to hydrogen- and helium-atmosphere
   candidates.}
   \label{fig:teffu}
\end{center}
\end{figure}

Another concern is the omission of the $u$-band photometry to estimate
our effective temperatures. Indeed, \citet[][see their Figures 4 and
  7]{bergeron2019} demonstrated that a much better agreement between
photometric and spectroscopic temperatures could be achieved if the
SDSS $u$ magnitude was combined with the Pan-STARRS $grizy$
photometry. To explore this effect, we compare in Figure
\ref{fig:teffu} the difference between effective temperatures obtained
by fitting Pan-STARRS $grizy$ photometry alone and the values obtained
by also including the $u$ magnitude (SDSS or CFIS) for objects within
the ZZ Ceti region. In this figure, different colors are used to
distinguish hydrogen- and helium-atmosphere candidates. Our results
indicate that for hydrogen-atmosphere white dwarfs in the range of
temperature of interest for our survey, the use of additional $u$-band
photometry has little effect on the estimated photometric
temperatures, with no systematic offset observed, and a standard
deviation of only 1.2\%.

The photometric instability strip displayed in the bottom panel of
Figure \ref{fig:fullsamp} can now be used to define a region that
contains 286 objects. From this list, we remove all known ZZ Ceti
pulsators taken from the compilation of \citet{corsico2019} as well as
recent discoveries \citep{romero2019}; these are indicated by magenta
symbols in the bottom panel of Figure
\ref{fig:fullsamp}. Incidentally, the location of these known
variables are perfectly well bracketed by our empirical photometric
instability strip, giving us confidence in our overall procedure.

Known helium-atmosphere white dwarfs --- cyan symbols in the bottom
panel of Figure \ref{fig:fullsamp} --- are also removed by comparing
our list against the MWDD and SIMBAD. Candidates with $u$ magnitudes
indicating a helium-rich atmosphere, through our fitting procedure
mentioned above, are also removed. While in principle this procedure
could be used to exclude all the remaining unidentified
helium-atmosphere candidates, $u$ magnitudes are only available for
less than half of the objects in our sample. 
Among the remaining candidates with available $u$ magnitudes,
about 26\% were removed through the fitting procedure, and so we expect 
a similar proportion of helium-atmosphere white dwarfs to contaminate our 
list of ZZ Ceti candidates with neither a $u$-band measurement nor known 
spectral information. The SDSS is the largest source of $u$ magnitudes
in our sample, but unfortunately, it does not cover as much sky as the
{\it Gaia} survey. The CFIS survey, currently under way\footnote{See
  \url{http://www.cfht.hawaii.edu/en/science/SAC/reports/SAC\_report\_November19.php}.},
should eventually provide $u$-band photometry for additional targets
in our sample. While its sky coverage mostly overlaps with SDSS, the
photometry will be approximately 3 magnitudes deeper than SDSS for a
given measurement uncertainty \citep{ibata2017}. The CFIS $u$
magnitudes have been consistent with our model predictions so far, as
displayed in Figure \ref{fig:fitphoto}.

Finally, objects in the southern hemisphere ($\delta<-10^\circ$) are
also excluded from our target list due to the location of the
Mont-M\'egantic Observatory, where our high-speed photometric
observations were secured. At the end, our final sample contains 173
ZZ Ceti candidates, out of which 80 are confirmed to be hydrogen-rich
through $u$-band photometry.

\section{Data Acquisition and Reduction}\label{sec:data}

We obtained time series photometry using the PESTO camera on the 1.6~m
telescope at the Mont-M\'egantic Observatory (Qu\'ebec). Our survey
spanned over 68 nights from 2018 July to 2020 August, using a mix of
classical and queue observing. We used a 10-second exposure time for
most observations, occasionally increasing to 30 seconds for fainter
objects. We initially used a $g'$ filter\footnote{See
  \url{http://omm-astro.ca/obs/instruments_www/pesto_dir/}.} but
eventually switched to using no filter to maximize the target flux and
signal-to-noise ratio. For an exposure time of 10 seconds, we achieved
a typical photometric precision of 2.6\% for objects with {\it Gaia}
magnitudes $15.5< G <16.5$, and 4.7\% for objects between $16.5< G
<17.5$. Our journal of observations is presented in
Table \ref{tab:journal}.

PESTO is a visible-light camera equipped with a 1024$\times$1024
pixels frame-transfer electron-multiplying (EM) CCD system from
N\"uv\"u Cameras. The pixel scale of $0.466''$ offers a $7.95'
\times7.95'$ field of view that allowed us to observe many neighboring
objects simultaneously, providing a better selection of comparison
stars for the data reduction. We operated the detector in conventional
mode, i.e., not using electron multiplication. The frame-transfer
operation of the CCD provides an observing efficiency near 100\%. The
camera is equipped with a time server based on Global Positioning
System for accurate timing of each exposure.

Our initial observational strategy was to observe every candidate for
one hour each, then, if pulsations were detected, to observe again for
an additional 4 hours. However due to the often varying and
unpredictable meteorological conditions at Mont-M\'egantic, such
4~h-long observations were often disrupted and difficult to
complete. Additionally, a single hour of initial observation was found
to be inadequate to detect long-period pulsators, which are expected
to have periods of up to 2000 seconds. Thus, about one year into the
survey, we decided to fix all of our observations to 2 hours per
candidate, aiming to maximize the quality of the data as well as the
number of candidates observed.

\afterpage{
\startlongtable
\begin{deluxetable}{clcccc}
\tablecaption{Journal of Observations\label{tab:journal}}
\tabletypesize{\footnotesize}

\tablehead{\colhead{Date at start} & \colhead{Gaia Source ID} & \colhead{Duration}& \colhead{No. of} & \colhead{Exp.} & \colhead{Filter} \\ 
\colhead{(UT)} & \colhead{} & \colhead{(h)}& \colhead{Images} & \colhead{(s)} & \colhead{} } 

\startdata
2020-08-13 04:01:06 & 2863526233218817024 & 1.5 & 361 & 15 & None\\ 
2020-08-13 05:40:04 & 2779284538516313600 & 1.3 & 451 & 10 & None\\ 
2020-08-13 07:01:05 & 2789405753503977472 & 1.5 & 361 & 15 & None\\ 
2020-08-11 02:35:41 & 2867203584218146944 & 1.0 & 241 & 15 & None\\ 
2020-08-08 02:11:13 & 4503347770490390016 & 1.5 & 361 & 15 & None\\ 
2020-08-08 03:53:26 & 1815614965310875520 & 1.5 & 361 & 15 & None\\ 
2020-08-08 05:28:25 & 1930609656643838080 & 1.5 & 361 & 15 & None\\ 
2020-08-07 03:21:24 & 4298401105174809984 & 1.9 & 451 & 15 & None\\ 
2020-08-07 05:08:55 & 1980205739970324224 & 1.7 & 408 & 15 & None\\ 
2020-08-07 06:53:00 & 1993426577008368640 & 1.6 & 381 & 15 & None\\ 
2020-07-29 03:56:58 & 4539136259802013952 & 1.3 & 451 & 10 & None\\ 
2020-07-22 03:15:46 & 2292229788249205760 & 1.6 & 559 & 10 & None\\ 
2020-07-16 02:56:53 & 2092086476924423808 & 2.2 & 522 & 15 & None\\ 
2020-07-16 05:22:28 & 2063435712171048704 & 1.3 & 451 & 10 & None\\ 
2020-07-10 03:38:09 & 1353302001211658368 & 1.6 & 381 & 15 & None\\ 
2020-07-07 05:59:56 & 2127591833389528064 & 2.0 & 484 & 15 & None\\ 
2020-06-20 06:39:37 & 2163226700308494080 & 1.3 & 313 & 15 & None\\ 
2020-06-19 04:49:00 & 1968901145520461568 & 1.6 & 376 & 15 & None\\ 
2020-06-19 03:22:53 & 1411867767238390912 & 1.3 & 451 & 10 & None\\ 
2020-06-19 06:50:59 & 2220815923910913920 & 1.3 & 451 & 10 & None\\ 
2020-06-18 02:34:25 & 1353355434900703616 & 1.3 & 451 & 10 & None\\ 
2020-06-17 03:23:41 & 575585919005741184 & 2.0 & 241 & 30 & None\\ 
2020-06-17 05:41:34 & 1845487489350432128 & 2.0 & 241 & 30 & None\\ 
2020-06-16 06:36:04 & 1344618951728016512 & 1.3 & 451 & 10 & None\\ 
2020-06-16 01:51:35 & 575585919005741184 & 2.3 & 271 & 30 & None\\ 
2020-06-12 01:47:26 & 2114985726416563072 & 2.3 & 278 & 30 & None\\ 
2020-06-06 01:33:41 & 1411867767238390912 & 1.6 & 566 & 10 & None\\ 
2020-03-15 23:43:00 & 3169486960220617088 & 1.9 & 700 & 10 & None\\ 
2020-03-16 08:10:55 & 1317275544951049472 & 2.0 & 717 & 10 & None\\ 
2020-02-15 06:48:28 & 3626525219143701120 & 2.0 & 721 & 10 & None\\ 
2020-01-31 07:10:47 & 642549544391197440 & 2.0 & 721 & 10 & None\\ 
2020-01-31 09:19:31 & 1587611884756030720 & 2.0 & 721 & 10 & None\\ 
2020-01-25 08:28:33 & 1456920737222542208 & 2.0 & 721 & 10 & None\\ 
2020-01-25 06:22:15 & 836410319296579712 & 2.0 & 721 & 10 & None\\ 
2019-11-24 02:38:19 & 3249740657527506048 & 2.2 & 803 & 10 & None\\ 
2019-11-17 06:53:55 & 63054590968017408 & 2.2 & 780 & 10 & None\\ 
2019-11-17 09:11:16 & 283096760659311744 & 1.9 & 667 & 10 & None\\ 
2019-10-22 00:14:38 & 2766498012855959424 & 2.0 & 721 & 10 & None\\ 
2019-10-20 07:50:29 & 3458597083113101952 & 2.0 & 721 & 10 & None\\ 
2019-10-19 23:25:10 & 4250461749665556224 & 2.0 & 721 & 10 & None\\ 
2019-10-20 01:31:38 & 2826770319713589888 & 2.0 & 721 & 10 & None\\ 
2019-10-14 07:49:56 & 3224908977688888064 & 2.4 & 878 & 10 & None\\ 
2019-10-09 05:45:12 & 302143768088623488 & 2.0 & 721 & 10 & None\\ 
2019-10-08 23:12:52 & 2177744858009335552 & 2.0 & 721 & 10 & None\\ 
2019-10-09 03:33:55 & 2844933221011789952 & 2.0 & 721 & 10 & None\\ 
2019-10-09 07:51:47 & 258439731372229120 & 2.0 & 721 & 10 & None\\ 
2019-10-06 04:15:49 & 192275966334956672 & 2.0 & 721 & 10 & None\\ 
2019-10-06 06:25:15 & 462506821746606464 & 2.0 & 721 & 10 & None\\ 
2019-10-05 23:05:55 & 2155960371551164416 & 2.0 & 721 & 10 & None\\ 
2019-10-05 02:38:08 & 1998740551069600128 & 2.0 & 721 & 10 & None\\ 
2019-10-04 23:58:34 & 2083300584444566016 & 2.5 & 902 & 10 & None\\ 
2019-10-05 04:42:33 & 377231345590861824 & 2.0 & 721 & 10 & None\\ 
2019-09-30 03:50:27 & 2746936704565640064 & 2.1 & 742 & 10 & None\\ 
2019-09-30 01:39:57 & 2811321837744375936 & 2.0 & 717 & 10 & None\\ 
2019-09-20 04:33:43 & 387724053774415104 & 2.3 & 551 & 15 & None\\ 
2019-09-20 01:52:23 & 2083661675243196544 & 2.3 & 551 & 15 & None\\ 
2019-09-19 23:46:41 & 1599685347062685184 & 1.9 & 551 & 12.5 & None\\ 
2019-09-19 02:41:07 & 2159171323461157120 & 3.1 & 551 & 20 & None\\ 
2019-09-13 03:59:20 & 135715232773818368 & 1.9 & 551 & 12.5 & None\\ 
2019-09-06 02:05:58 & 1631796309274519040 & 2.2 & 600 & 13 & None\\ 
2019-08-26 00:34:30 & 4555079659441944960 & 2.3 & 551 & 15 & None\\ 
2019-08-26 02:57:38 & 1842670231320998016 & 1.5 & 551 & 10 & None\\ 
2019-08-24 00:53:59 & 2263690864438162944 & 2.3 & 551 & 15 & None\\ 
2019-08-06 01:17:31 & 4454017257893306496 & 2.5 & 604 & 15 & None\\ 
2019-08-06 03:58:34 & 2086392484163910656 & 2.1 & 600 & 12.5 & None\\ 
2019-08-05 07:12:28 & 1998740551069600128 & 1.6 & 560 & 10 & None\\ 
2019-08-03 05:20:47 & 1793328410074430464 & 3.3 & 537 & 22 & None\\ 
2019-08-02 01:48:56 & 1631796309274519040 & 2.4 & 551 & 16 & None\\ 
2019-07-27 06:43:57 & 302143768088623488 & 1.9 & 451 & 15 & None\\ 
2019-07-27 01:20:57 & 4555079659441944960 & 3.0 & 720 & 15 & None\\ 
2019-07-27 04:32:36 & 2263690864438162944 & 2.0 & 721 & 10 & None\\ 
2019-07-10 01:38:58 & 2055661546498684416 & 2.0 & 716 & 10 & None\\ 
2019-07-10 03:39:55 & 1793328410074430464 & 2.0 & 716 & 10 & None\\ 
2019-07-10 05:42:38 & 1913174219724912128 & 2.1 & 756 & 10 & None\\ 
2019-07-08 01:30:48 & 4447022061837071744 & 2.2 & 809 & 10 & $g'$\\ 
2019-07-03 01:38:29 & 2159171323461157120 & 2.2 & 787 & 10 & None\\ 
2019-07-03 04:02:39 & 2086392484163910656 & 2.0 & 729 & 10 & None\\ 
2019-07-03 06:08:16 & 2263690864438162944 & 2.0 & 711 & 10 & None\\ 
2019-06-25 06:45:20 & 4337833650892408448 & 2.1 & 769 & 10 & None\\ 
2019-06-25 10:01:45 & 4217910669267424512 & 2.2 & 794 & 10 & None\\ 
2019-06-24 10:19:39 & 4498531123585093120 & 2.1 & 750 & 10 & None\\ 
2019-06-23 10:12:08 & 4491980748701631616 & 2.1 & 758 & 10 & None\\ 
2019-06-18 07:07:21 & 1543370904111505408 & 2.1 & 743 & 10 & $g'$\\ 
2019-06-12 10:14:50 & 2265100885021724032 & 0.8 & 296 & 10 & $g'$\\ 
2019-06-12 11:16:24 & 2263690864438162944 & 0.8 & 304 & 10 & $g'$\\ 
2019-06-12 08:08:27 & 2083661675243196544 & 0.8 & 273 & 10 & $g'$\\ 
2019-05-28 09:40:15 & 4337833650892408448 & 0.8 & 298 & 10 & $g'$\\ 
2019-05-28 10:43:04 & 4336571785203401472 & 0.8 & 299 & 10 & $g'$\\ 
2019-05-28 11:45:06 & 4498531123585093120 & 0.8 & 304 & 10 & $g'$\\ 
2019-04-05 00:36:20 & 672816969200760064 & 2.0 & 1464 & 5 & $g'$\\ 
2019-04-05 03:17:39 & 1042926292644833024 & 1.0 & 357 & 10 & $g'$\\ 
2019-04-02 05:41:07 & 4570546317703725312 & 4.0 & 1438 & 10 & $g'$\\ 
2019-03-30 07:31:02 & 4349734833473621248 & 1.0 & 372 & 10 & $g'$\\ 
2019-03-28 05:36:35 & 4454017257893306496 & 1.2 & 447 & 10 & $g'$\\ 
2019-03-28 07:14:46 & 1304081783374935680 & 1.2 & 448 & 10 & $g'$\\ 
2019-03-28 08:18:40 & 4555079659441944960 & 1.3 & 459 & 10 & $g'$\\ 
2019-03-24 02:30:22 & 1042926292644833024 & 2.2 & 779 & 10 & $g'$\\ 
2019-03-24 07:32:59 & 4555079659441944960 & 2.5 & 892 & 10 & $g'$\\ 
2019-03-18 23:38:39 & 53716846734195328 & 2.4 & 864 & 10 & $g'$\\ 
2019-03-19 05:41:01 & 3719371829283488768 & 2.0 & 731 & 10 & $g'$\\ 
2019-03-13 01:34:39 & 1042926292644833024 & 1.2 & 425 & 10 & $g'$\\ 
2019-03-01 23:14:43 & 377231139432432384 & 1.0 & 357 & 10 & $g'$\\ 
2019-03-02 04:33:50 & 672816969200760064 & 1.0 & 350 & 10 & $g'$\\ 
2019-03-02 02:26:12 & 3080844435869554176 & 1.0 & 374 & 10 & $g'$\\ 
2019-03-02 03:30:07 & 3150770626615542784 & 1.0 & 370 & 10 & $g'$\\ 
2019-03-02 06:38:26 & 3937174946624964224 & 1.0 & 366 & 10 & $g'$\\ 
2019-03-02 07:40:43 & 3719371829283488768 & 1.0 & 357 & 10 & $g'$\\ 
2019-03-02 08:50:12 & 4454017257893306496 & 0.9 & 331 & 10 & $g'$\\ 
2019-02-28 23:41:44 & 3400048535611299456 & 4.0 & 1441 & 10 & $g'$\\ 
2019-02-28 01:25:33 & 1682022481467013504 & 1.0 & 361 & 10 & $g'$\\ 
2019-02-28 07:20:00 & 1456920737222542208 & 1.0 & 361 & 10 & $g'$\\ 
2019-02-28 08:27:13 & 1316268323580640256 & 1.0 & 361 & 10 & $g'$\\ 
2019-02-28 09:35:35 & 1304274094830734720 & 1.0 & 361 & 10 & $g'$\\ 
2019-02-23 23:21:29 & 412839403319209600 & 1.0 & 361 & 10 & $g'$\\ 
2019-02-23 06:19:59 & 1543370904111505408 & 1.0 & 361 & 10 & $g'$\\ 
2019-02-23 08:47:24 & 1566530913957066240 & 1.0 & 361 & 10 & $g'$\\ 
2019-02-19 23:06:41 & 377231139432432384 & 1.0 & 377 & 10 & $g'$\\ 
2019-02-20 00:29:36 & 3400048535611299456 & 4.0 & 1444 & 10 & $g'$\\ 
2019-02-17 23:48:16 & 436085007572835072 & 1.1 & 402 & 10 & $g'$\\ 
2019-02-11 02:52:09 & 647899806626643200 & 1.0 & 361 & 10 & $g'$\\ 
2019-01-27 01:42:12 & 3181589319065856384 & 1.0 & 361 & 10 & $g'$\\ 
2019-01-27 02:54:23 & 3439162768415866112 & 1.0 & 361 & 10 & $g'$\\ 
2019-01-27 04:01:17 & 945007674022721280 & 1.0 & 361 & 10 & $g'$\\ 
2019-01-27 05:07:52 & 1087442842689746048 & 1.0 & 361 & 10 & $g'$\\ 
2019-01-14 05:14:56 & 184735992329821312 & 1.0 & 361 & 10 & $g'$\\ 
2019-01-14 08:28:49 & 1114813977776610944 & 1.0 & 361 & 10 & $g'$\\ 
2019-01-14 09:58:20 & 791138993175412480 & 0.7 & 261 & 10 & $g'$\\ 
2018-12-13 10:01:58 & 983538336734107392 & 1.2 & 450 & 10 & $g'$\\ 
2018-11-12 06:43:21 & 3447991090873280000 & 1.0 & 365 & 10 & $g'$\\ 
2018-11-12 07:55:02 & 3400048535611299456 & 1.0 & 368 & 10 & $g'$\\ 
2018-09-24 23:23:15 & 1897597369775277568 & 4.1 & 1481 & 10 & $g'$\\ 
2018-09-23 02:07:19 & 1998740551069600128 & 1.0 & 361 & 10 & $g'$\\ 
2018-09-15 05:37:38 & 2778812676229535616 & 1.0 & 365 & 10 & $g'$\\ 
2018-09-15 04:06:47 & 387724053774415104 & 1.0 & 364 & 10 & $g'$\\ 
2018-09-15 06:53:36 & 415684119076509056 & 1.3 & 464 & 10 & $g'$\\ 
2018-09-10 00:01:22 & 4570546317703725312 & 1.0 & 361 & 10 & $g'$\\ 
2018-09-10 02:29:11 & 1897597369775277568 & 1.0 & 361 & 10 & $g'$\\ 
2018-09-10 01:13:18 & 1835056216381670272 & 1.0 & 361 & 10 & $g'$\\ 
2018-08-25 07:50:24 & 2647884790098989568 & 1.3 & 472 & 10 & $g'$\\ 
2018-08-24 00:39:53 & 2114811453822316160 & 4.5 & 1627 & 10 & $g'$\\ 
2018-08-21 07:25:21 & 2826770319713589888 & 1.6 & 589 & 10 & $g'$\\ 
2018-08-20 07:43:31 & 2844933221011789952 & 0.6 & 199 & 10 & $g'$\\ 
2018-08-20 06:44:37 & 1913174219724912128 & 0.9 & 322 & 10 & $g'$\\ 
2018-08-19 01:44:57 & 4281190419601308672 & 1.0 & 364 & 10 & $g'$\\ 
2018-08-19 02:48:21 & 4321498378443922816 & 0.7 & 252 & 10 & $g'$\\ 
2018-08-17 03:58:48 & 2055661546498684416 & 1.0 & 368 & 10 & $g'$\\ 
2018-08-01 02:23:07 & 2240031951187372928 & 0.9 & 341 & 10 & $g'$\\ 
2018-07-31 01:33:09 & 1631796309274519040 & 1.0 & 363 & 10 & $g'$\\ 
2018-07-31 06:55:15 & 1995097319287822080 & 0.8 & 286 & 10 & $g'$\\ 
2018-07-31 05:47:10 & 2083300584444566016 & 0.8 & 296 & 10 & $g'$\\ 
2018-07-30 03:50:55 & 2114811453822316160 & 1.0 & 356 & 10 & $g'$\\ 
2018-07-30 01:35:26 & 2159171323461157120 & 1.0 & 354 & 10 & $g'$\\ 
\enddata
\end{deluxetable}} 

We reduced the data using custom Python scripts and following standard
procedures. The raw data frames were first bias and dark subtracted
and flat-field corrected. Then, for each calibrated frame, we used the
\texttt{Astropy} \citep{astropy2013} and \texttt{Photutils}
\citep{Bradley2019} Python packages to perform circular aperture
photometry to extract the sky-subtracted flux of the target and a
number of neighboring stars. For a typical point spread function (PSF)
of 5.3 pixels FWHM, we used an aperture radius of 6 pixels and a sky
annulus inner and outer radius of 18 and 23 pixels, respectively. The
resulting light curves were then normalized to their median value. To
correct for atmospheric and instrumental effects, we divided the
target light curve by the median light curve of two or more comparison
stars, prioritizing those with similar magnitudes and colors. We also
verified that the comparison stars were photometrically constant by
looking at their own calibrated light curve. After this first
calibration, the light curves were airmass detrended using a second or
third order polynomial, and the previous calibration process was
repeated once. Finally, we computed a Lomb-Scargle periodogram of the
candidate light curve using the custom implementation of
\citet{townsend2010} for unevenly-spaced data, as some light curves
were fragmented due to meteorological conditions.

\clearpage
\section{Results}\label{sec:res}

\subsection{New Variables and Non-variables}

High-speed photometric observations were secured for 90 ZZ Ceti
candidates, out of which 38 were clearly variable, 5 showed possible
weak periodic signals (see below), and 47 were not observed to vary
(NOV). We also observed 18 additional objects located above the hot
edge of the photometric instability strip, which were part of our
prior selection of candidates based on the spectroscopic instability
strip from \citet{green2015}. Although none of these turned out to be
variable, they remain valuable objects to determine the exact location
of the blue edge of the strip.

The new ZZ Ceti white dwarfs and possible pulsators are presented in
Table \ref{tab:var} along with the WD ID\footnote{The WD ID numbers
  JXXXX+YYYY assigned here are based on the {\it Gaia} J2015.5
  coordinates.}, {\it Gaia} ID, right ascension, declination,
effective temperature, stellar mass, {\it Gaia} $G$ magnitude, SDSS or
CFIS $u$ magnitude, and literature identifying the
object as a DA, if available; the possible pulsators are denoted with
a colon at the end of the WD ID. Note that the $u$-band photometry is
included in the photometric fits used to measure the physical
parameters given here, and in every result discussed henceforth. Also
reported in Table \ref{tab:var} are the dominant periods and
amplitudes, which will be discussed later in Section \ref{sec:puls}.

Light curves for every new ZZ Ceti star and possible pulsator in our
sample are presented in Figure \ref{fig:lcfft}. A quick examination of
these results reveal a rich variety of short- and long-period
pulsators. In general, the long-period variables tend to have the
largest amplitudes, but this is not always the case (see, e.g.,
J1058$+$5132). We also find triangular-shaped pulsations, indicative
of the presence of harmonics, as well as a few cases of beats, which
reveal the presence of closely-spaced oscillation modes. The
variability of most objects displayed in Figure \ref{fig:lcfft} can be
clearly assessed based on the light curves alone, but some require a
more quantitative inspection. To this end, the Lomb-Scargle
periodograms are shown next to each light curve in Figure
\ref{fig:lcfft}, covering a frequency spectrum ranging from 0.01~mHz
up to 10.5~mHz. The region covering 10.5~mHz up to the Nyquist
frequency (50~mHz for a 10~s sampling time) is always consistent with
noise and is therefore not shown.

To estimate the chance that the detected signals are
  real, we calculate the FAP (False Alarm Probability) using the
  bootstrap method described in \citet{vanderplas2018}. If the dominant
  periodic signal can be verified by eye and/or has a FAP smaller than
  0.1\%, we then consider the object as a new variable white dwarf.
  Objects that fail this criterion but that nevertheless show a 
  periodic signal with an amplitude larger than 5 times the mean of 
  the entire periodogram are classified as possible pulsators.
  The two quantities used for classification are included in Table 
  \ref{tab:var}, and possible pulsators are identified with a colon 
  in both Table \ref{tab:var} and Figure \ref{fig:lcfft}.
These objects mostly correspond to candidates located close to
the instability strip edges, which are expected to show small
amplitudes, thus making their variability more difficult to
detect. Some of these signals might be buried by the noise of
sub-optimal observing conditions, while some might simply be near or
below our observational limits. We further discuss our possible
pulsators in Section \ref{sec:puls}.

\begin{longrotatetable}
\startlongtable
\begin{deluxetable}{lrccccccrcccl}
\tablecaption{New ZZ Ceti White Dwarfs and Possible Pulsators Properties\label{tab:var}}
\tabletypesize{\scriptsize}

\tablehead{\colhead{WD} & \colhead{Gaia DR2 Source} & \colhead{R.A.} & \colhead{Dec.} & \colhead{P} & \colhead{Amp} & \colhead{5$\sigma$} & \colhead{FAP} & \colhead{$\Te$} & \colhead{$M$} & \colhead{$G$} & \colhead{$u$} & \colhead{DA Classification}\\ 
\colhead{} & \colhead{} & \colhead{(J2015.5)} & \colhead{(J2015.5)} &  \colhead{(s)} & \colhead{(\%)} & \colhead{(\%)} & \colhead{(\%)} &\colhead{(K)} & \colhead{(\msun)} & \colhead{} & \colhead{} & \colhead{}} 

\startdata
J0013$+$3246 & 2863526233218817024 & 00:13:19.80 & $+$32:46:12.96 & 1459 & 0.2 & 0.09 & $<$0.1 & 10311 $\pm$ 54 & 0.538 $\pm$ 0.009 & 16.7 & 17.1 \tablenotemark{\scriptsize a} & \citet{kilic2020} \\
J0039$+$1318 & 2779284538516313600 & 00:39:29.25 & $+$13:18:05.93 & 1579 & 0.3 & 0.13 & $<$0.1 & 10740 $\pm$ 94 & 0.591 $\pm$ 0.010 & 16.4 & 16.8 \tablenotemark{\scriptsize a} & \citet{kilic2020}  \\
J0049$+$2027: & 2789405753503977472 & 00:49:29.44 & $+$20:27:11.21 & 1102 & 0.2 & 0.08 & 0.4    & 10524 $\pm$ 73 & 0.586 $\pm$ 0.013 & 17.0 & 17.4 \tablenotemark{\scriptsize a} & -  \\ 
J0139$+$2900: & 302143768088623488 & 01:39:14.43 & $+$29:00:57.21 & 143 & 0.2  & 0.08 & 0.2    & 11625 $\pm$ 76 & 0.686 $\pm$ 0.008 & 16.4 & 16.7 \tablenotemark{\scriptsize a} & \citet{zhang2013}  \\ 
J0204$+$8713 & 575585919005741184 & 02:04:31.02 & $+$87:13:32.84 & 330 & 0.8   & 0.30 & 0.2  & 11131 $\pm$ 206 & 1.049 $\pm$ 0.015 & 17.8 &  - & -  \\ 
J0302$+$4800 & 436085007572835072 & 03:02:11.40 & $+$48:00:13.58 & 377 & 8.1   & 2.61 & $<$0.1 & 11551 $\pm$ 60 & 0.614 $\pm$ 0.006 & 16.3 &  - & -  \\ 
J0324$+$6020 & 462506821746606464 & 03:24:38.66 & $+$60:20:55.88 & 900 & 1.4   & 0.24 & $<$0.1 & 10826 $\pm$ 76 & 0.611 $\pm$ 0.008 & 16.1 &  - & -  \\ 
J0433$+$4850 & 258439731372229120 & 04:33:50.99 & $+$48:50:39.18 & 1029 & 4.9  & 1.50 & $<$0.1 & 10952 $\pm$ 121 & 0.57 $\pm$ 0.009 & 15.9 &  - & -  \\ 
J0448$-$1053 & 3181589319065856384 & 04:48:32.07 & $-$10:53:50.09 & 521 & 14.7 & 2.94 & $<$0.1 & 11993 $\pm$ 108 & 0.941 $\pm$ 0.006 & 16.3 &  - & -  \\ 
J0451$-$0333 & 3224908977688888064 & 04:51:32.19 & $-$03:33:08.43 & 908 & 22.4 & 3.67 & $<$0.1 & 10927 $\pm$ 79 & 0.598 $\pm$ 0.008 & 16.1 & 16.5 \tablenotemark{\scriptsize a} & \citet{kilic2020}  \\
J0546$+$2055 & 3400048535611299456 & 05:46:02.09 & $+$20:55:58.34 & 196 & 0.8  & 0.26 & $<$0.1 &  11632 $\pm$ 62 & 0.571 $\pm$ 0.008 & 16.4 & 16.8 \tablenotemark{\scriptsize a} & \citet{kilic2020}  \\
J0551$+$4135 & 192275966334956672 & 05:51:34.61 & $+$41:35:31.09 & 809 & 0.4   & 0.10 & $<$0.1 & 12513 $\pm$ 117 & 1.127 $\pm$ 0.005 & 16.4 &  - & -  \\ 
J0557$+$4034 & 3458597083113101952 & 05:57:17.68 & $+$40:34:36.76 & 256 & 0.3  & 0.07 & $<$0.1 & 11593 $\pm$ 144 & 0.537 $\pm$ 0.012 & 16.4 &  - & -  \\ 
J0723$+$1617 & 3169486960220617088 & 07:23:00.20 & $+$16:17:04.80 & 491 & 10.8 & 1.54 & $<$0.1 & 11448 $\pm$ 104 & 0.793 $\pm$ 0.008 & 15.1 &  - & -  \\ 
J0737$+$5215: & 983538336734107392 & 07:37:19.29 & $+$52:15:06.32 & 256 & 0.7  & 0.41 & 5.6    & 11544 $\pm$ 105 & 0.576 $\pm$ 0.010 & 16.7 &  - & -  \\ 
J0856$+$6206 & 1042926292644833024 & 08:56:19.34 & $+$62:06:32.59 & 415 & 5.1  & 2.3 & $<$0.1 & 11855 $\pm$ 72 & 0.959 $\pm$ 0.007 & 17.0 & 17.2 \tablenotemark{\scriptsize a} & \citet{kilic2020}  \\ 
J0938$+$2758 & 647899806626643200 & 09:38:07.10 & $+$27:58:20.09 & 563 & 14.3  & 3.14 & $<$0.1 & 11419 $\pm$ 104 & 0.815 $\pm$ 0.015 & 17.1 & 17.4 \tablenotemark{\scriptsize a} & \citet{guo2015}  \\ 
J1004$+$2438 & 642549544391197440 & 10:04:12.46 & $+$24:38:49.45 & 783 & 5.8   & 0.81 & $<$0.1 & 10919 $\pm$ 66 & 0.589 $\pm$ 0.010 & 16.5 & 16.9 \tablenotemark{\scriptsize a} & \citet{limoges2015}  \\
J1058$+$5132 & 836410319296579712 & 10:58:38.58 & $+$51:32:38.18 & 880 & 1.0   & 0.18 & 1.2    & 10819 $\pm$ 58 & 0.569 $\pm$ 0.011 & 16.5 & 16.9 \tablenotemark{\scriptsize a} & -  \\ 
J1207$+$6855: & 1682022481467013504 & 12:07:46.11 & $+$68:55:55.70 & 102 & 0.7 & 0.63 & 42     & 12255 $\pm$ 96 & 0.761 $\pm$ 0.007 & 16.8 & 17.1 \tablenotemark{\scriptsize a} & \citet{kilic2020}  \\
J1250$-$1042: & 3626525219143701120 & 12:50:27.19 & $-$10:42:39.20 & 258 & 0.6 & 0.29 & 0.7 & 11257 $\pm$ 59 & 0.529 $\pm$ 0.010 & 16.5 &  - & -  \\ 
J1314$+$1732 & 3937174946624964224 & 13:14:26.80 & $+$17:32:08.62 & 257 & 12.1 & 4.09 & $<$0.1 & 11505 $\pm$ 109 & 0.592 $\pm$ 0.009 & 16.3 & 16.7 \tablenotemark{\scriptsize a} & \citet{andrews2015}  \\ 
J1352$+$3012 & 1456920737222542208 & 13:52:11.18 & $+$30:12:34.48 & 195 & 0.7  & 0.09 & $<$0.1 & 11585 $\pm$ 47 & 0.629 $\pm$ 0.006 & 16.1 & 16.4 \tablenotemark{\scriptsize a} & \citet{kilic2020}  \\ 
J1509$+$4546 & 1587611884756030720 & 15:09:45.35 & $+$45:46:24.41 & 814 & 4.9  & 0.67 & $<$0.1 & 11180 $\pm$ 71 & 0.639 $\pm$ 0.007 & 16.5 & 16.8 \tablenotemark{\scriptsize a} & \citet{kilic2020}  \\ 
J1718$+$2524 & 4570546317703725312 & 17:18:40.61 & $+$25:24:31.53 & 731 & 38.5 & 6.10 & $<$0.1 & 11351 $\pm$ 98 & 0.628 $\pm$ 0.008 & 16.1 & 16.5 \tablenotemark{\scriptsize a} & \citet{kilic2020}  \\
J1730$+$1052 & 4491980748701631616 & 17:30:42.89 & $+$10:52:45.48 & 261 & 3.7  & 0.38 & $<$0.1 & 11373 $\pm$ 127 & 0.572 $\pm$ 0.009 & 16.2 &  - & -  \\ 
J1757$+$1803 & 4503347770490390016 & 17:57:40.88 & $+$18:03:55.49 & 857 & 0.3  & 0.96 & $<$0.1 & 10377 $\pm$ 95 & 0.542 $\pm$ 0.012 & 16.6 &  - & -  \\ 
J1812$+$4321 & 2114811453822316160 & 18:12:22.75 & $+$43:21:08.24 & 355 & 2.5  & 0.40 & $<$0.1 & 12448 $\pm$ 103 & 0.917 $\pm$ 0.006 & 16.3 & 16.4 \tablenotemark{\scriptsize a} & \citet{kilic2020}  \\
J1813$+$6220 & 2159171323461157120 & 18:13:57.78 & $+$62:20:10.47 & 370 & 1.2  & 0.17 & $<$0.1 & 11539 $\pm$ 140 & 0.848 $\pm$ 0.013 & 17.3 &  - & -  \\ 
J1843$+$2740 & 4539136259802013952 & 18:43:35.64 & $+$27:40:25.45 & 968 & 0.5  & 0.09 & $<$0.1 & 10566 $\pm$ 57 & 0.603 $\pm$ 0.006 & 15.0 &  - & \citet{limoges2015}  \\ 
J1903$+$6035 & 2155960371551164416 & 19:03:19.56 & $+$60:35:52.65 & 726 & 10.3 & 1.53 & $<$0.1 & 10858 $\pm$ 63 & 0.624 $\pm$ 0.006 & 15.0 &  - & \citet{limoges2015}  \\ 
J1925$+$4641 & 2127591833389528064 & 19:25:05.05 & $+$46:41:04.33 & 844 & 0.6  & 0.14 & $<$0.1 & 10655 $\pm$ 121 & 0.619 $\pm$ 0.013 & 16.9 &  - & -  \\ 
J1928$+$6105 & 2240031951187372928 & 19:28:53.71 & $+$61:05:48.71 & 302 & 7.6  & 1.57 & $<$0.1 & 11253 $\pm$ 126 & 0.585 $\pm$ 0.009 & 16.4 &  - & -  \\ 
J2013$+$3413 & 2055661546498684416 & 20:13:43.42 & $+$34:13:56.88 & 549 & 4.6  & 1.31 & $<$0.1 & 11440 $\pm$ 118 & 0.854 $\pm$ 0.009 & 15.7 &  - & -  \\ 
J2013$+$0709 & 4250461749665556224 & 20:13:53.31 & $+$07:09:45.15 & 206 & 4.6  & 0.59 & $<$0.1 & 11645 $\pm$ 84 & 0.656 $\pm$ 0.009 & 16.5 & 16.8 \tablenotemark{\scriptsize a} & \citet{kilic2020}  \\ 
J2023$-$0620 & 4217910669267424512 & 20:23:18.61 & $-$06:20:15.63 & 497 & 8.4  & 1.01 & $<$0.1 & 11081 $\pm$ 94 & 0.606 $\pm$ 0.011 & 16.7 &  - & -  \\ 
J2150$+$3035 & 1897597369775277568 & 21:50:40.54 & $+$30:35:37.16 & 335 & 1.6  & 0.69 & $<$0.1 & 11429 $\pm$ 79 & 0.562 $\pm$ 0.007 & 16.0 &  - & -  \\ 
J2159$+$5102 & 1980205739970324224 & 21:59:17.26 & $+$51:02:56.42 & 1286 & 1.2 & 0.27 & $<$0.1 & 10936 $\pm$ 146 & 0.864 $\pm$ 0.015 & 17.1 &  - & -  \\ 
J2319$+$2728 & 2844933221011789952 & 23:19:36.27 & $+$27:28:58.17 & 277 & 1.4  & 0.39 & $<$0.1 & 10463 $\pm$ 92 & 0.505 $\pm$ 0.012 & 16.3 & 16.8 \tablenotemark{\scriptsize a} & -  \\ 
J2322$+$3605 & 1913174219724912128 & 23:22:15.56 & $+$36:05:44.05 & 363 & 5.0  & 0.70 & $<$0.1 & 11265 $\pm$ 39 & 0.585 $\pm$ 0.006 & 16.3 & 16.6 \tablenotemark{\scriptsize b} & -  \\ 
J2346$+$2200 & 2826770319713589888 & 23:46:33.67 & $+$22:00:42.63 & 1161 & 0.3 & 0.11 & $<$0.1 & 11078 $\pm$ 72 & 0.541 $\pm$ 0.009 & 16.5 & 16.8 \tablenotemark{\scriptsize a} & \citet{kilic2020}  \\
J2353$+$2928 & 2867203584218146944 & 23:53:18.31 & $+$29:28:08.87 & 545 & 4.7  & 0.85 & $<$0.1 & 11146 $\pm$ 72 & 0.812 $\pm$ 0.010 & 17.1 & 17.4 \tablenotemark{\scriptsize a} & -  \\ 
J2356$+$1143 & 2766498012855959424 & 23:56:37.43 & $+$11:43:35.92 & 252 & 0.5 & 0.20 & $<$0.1 & 11745 $\pm$ 80 & 0.665 $\pm$ 0.008 & 16.4 & 16.7 \tablenotemark{\scriptsize a} & \citet{kilic2020}  \\ 
\enddata
\tablenotetext{a}{SDSS photometry.}
\tablenotetext{b}{CFIS photometry.}
\end{deluxetable}
\end{longrotatetable} 

The difference in quality between filtered and unfiltered light curves
can be appreciated by comparing J0302$+$4800 and J0551$+$4135 in
Figure \ref{fig:lcfft}. Both have similar {\it Gaia} magnitudes and
seeing --- $G\sim16.33$ and 16.37, FWHM $\sim$5.8 and 6.0,
respectively --- but the first has been observed with the $g'$ filter,
while the latter has been observed in white light. The pulsations for
the object observed in white light are much more obvious, even though
it is a shorter-period and smaller-amplitude pulsator than the object
observed with a filter.

\begin{figure}
\figurenum{7}
\begin{center}
 \includegraphics[width=0.94\columnwidth,clip=True,trim=0 0 0 0.66in]{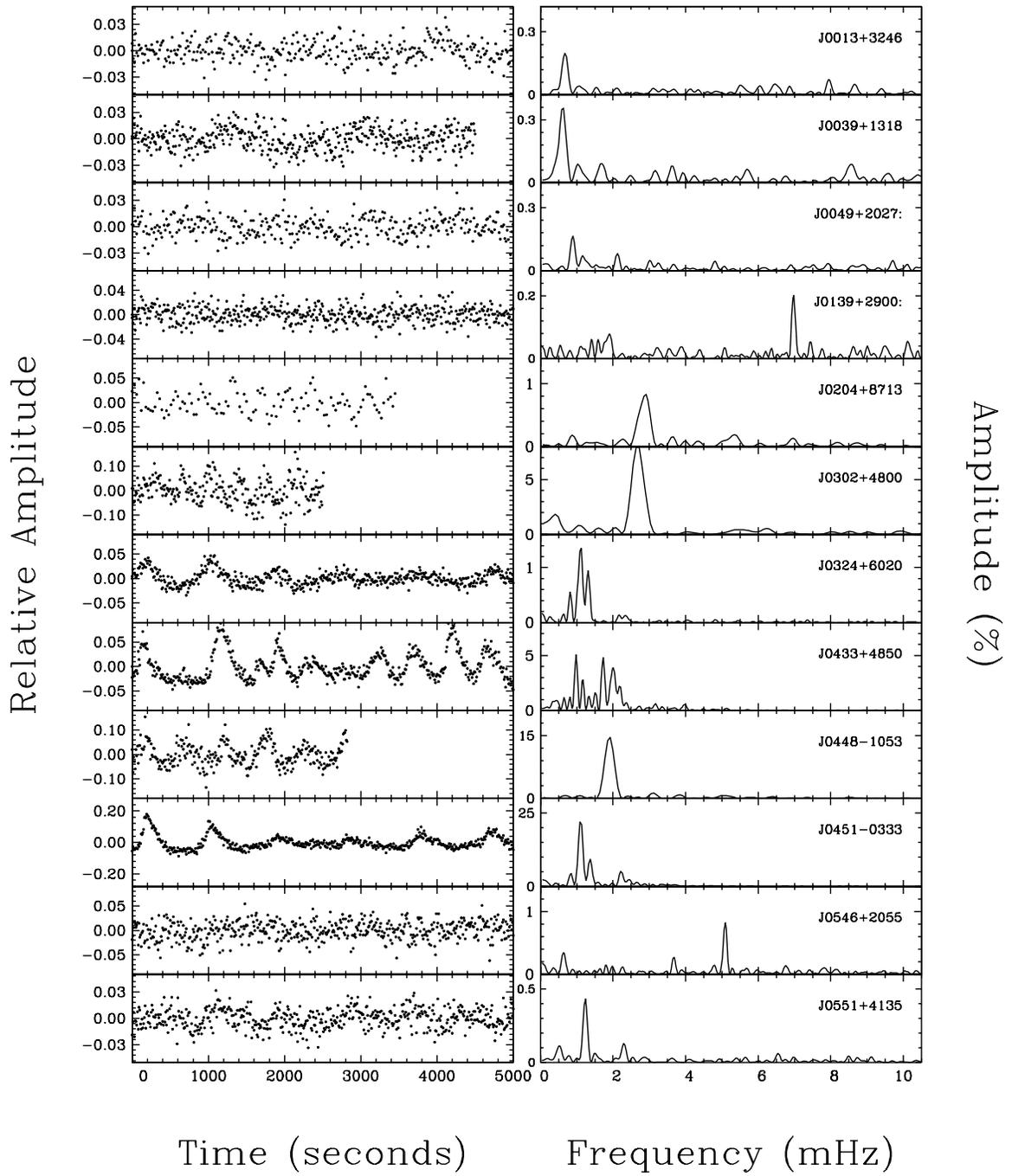}
 \caption{Light curves and Lomb-Scargle periodograms for the newly
   discovered ZZ Ceti white dwarfs and possible pulsators. The
   periodogram amplitude is expressed in terms of the percentage
   variations about the mean brightness of the star.}
   \label{fig:lcfft}
\end{center}
\end{figure}

\begin{figure}
\figurenum{7}
\begin{center}
 \includegraphics[width=0.94\columnwidth,clip=True,trim=0 0 0 0.66in]{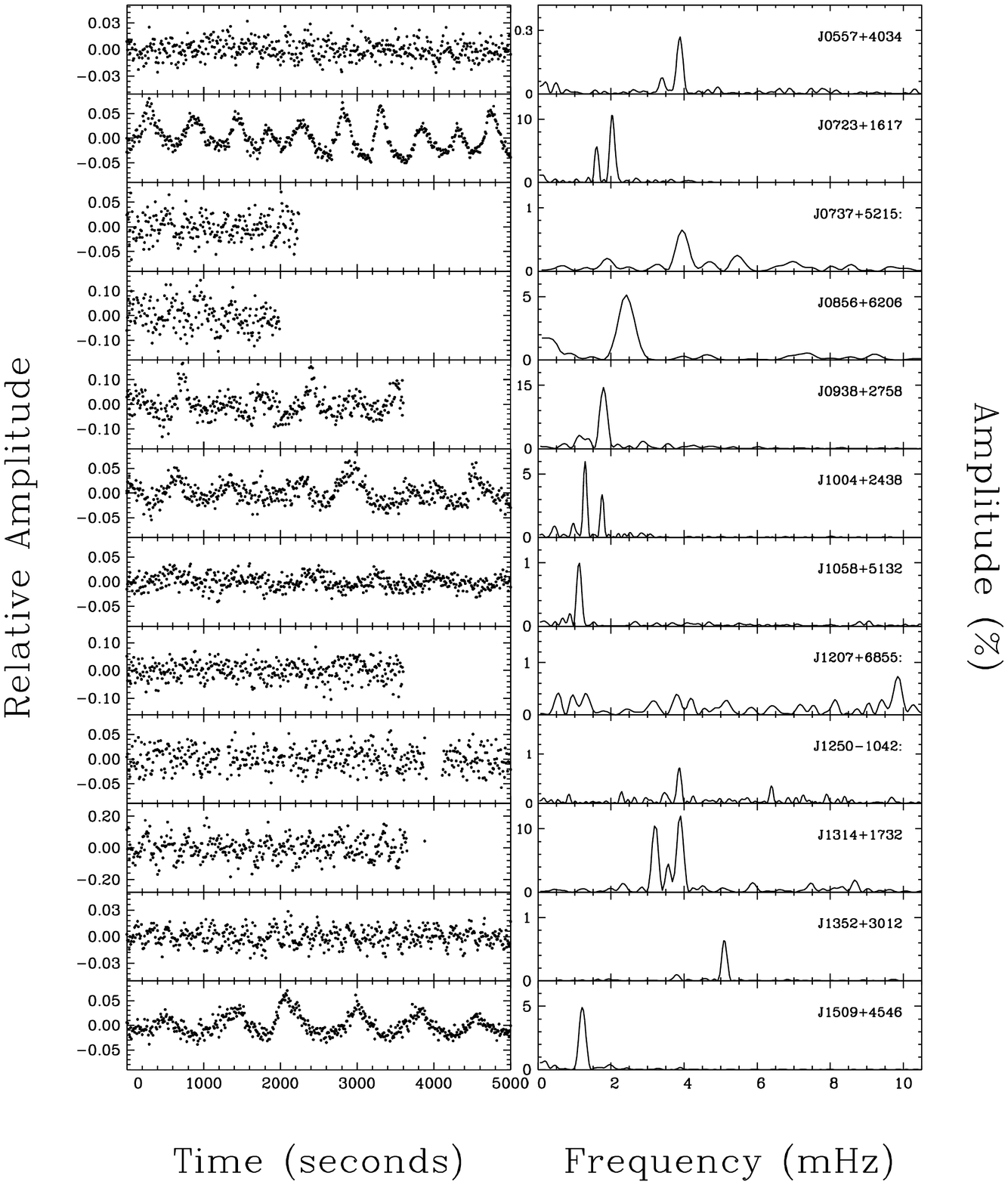}
 \caption{(Continued)}
\end{center}
\end{figure}

\begin{figure}
\figurenum{7}
\begin{center}
 \includegraphics[width=0.94\columnwidth,clip=True,trim=0 0 0 0.66in]{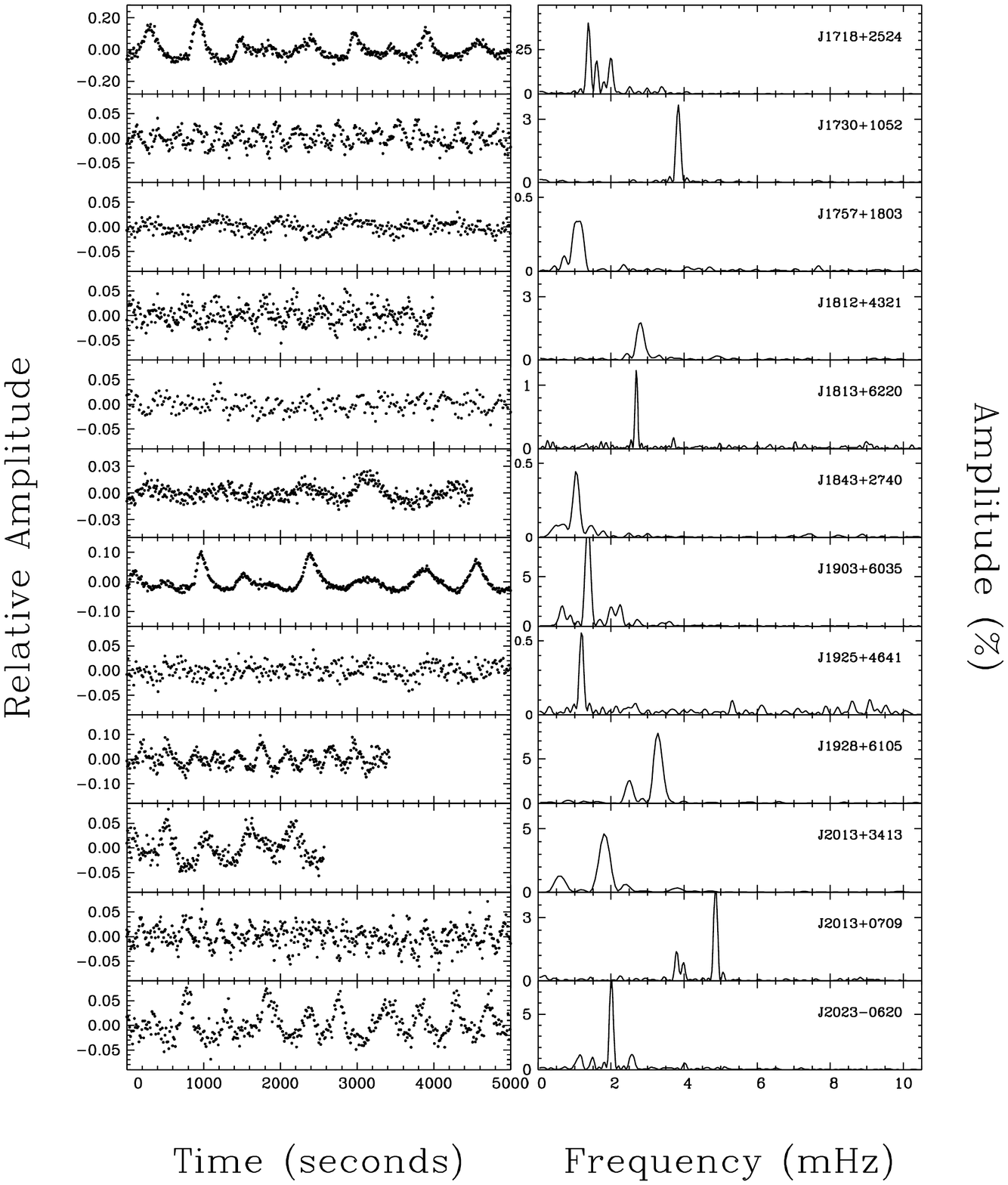}
 \caption{(Continued)}
\end{center}
\end{figure}

\begin{figure}
\figurenum{7}
\begin{center}
 \includegraphics[width=0.94\columnwidth,clip=True,trim=0 0 0 4.1in]{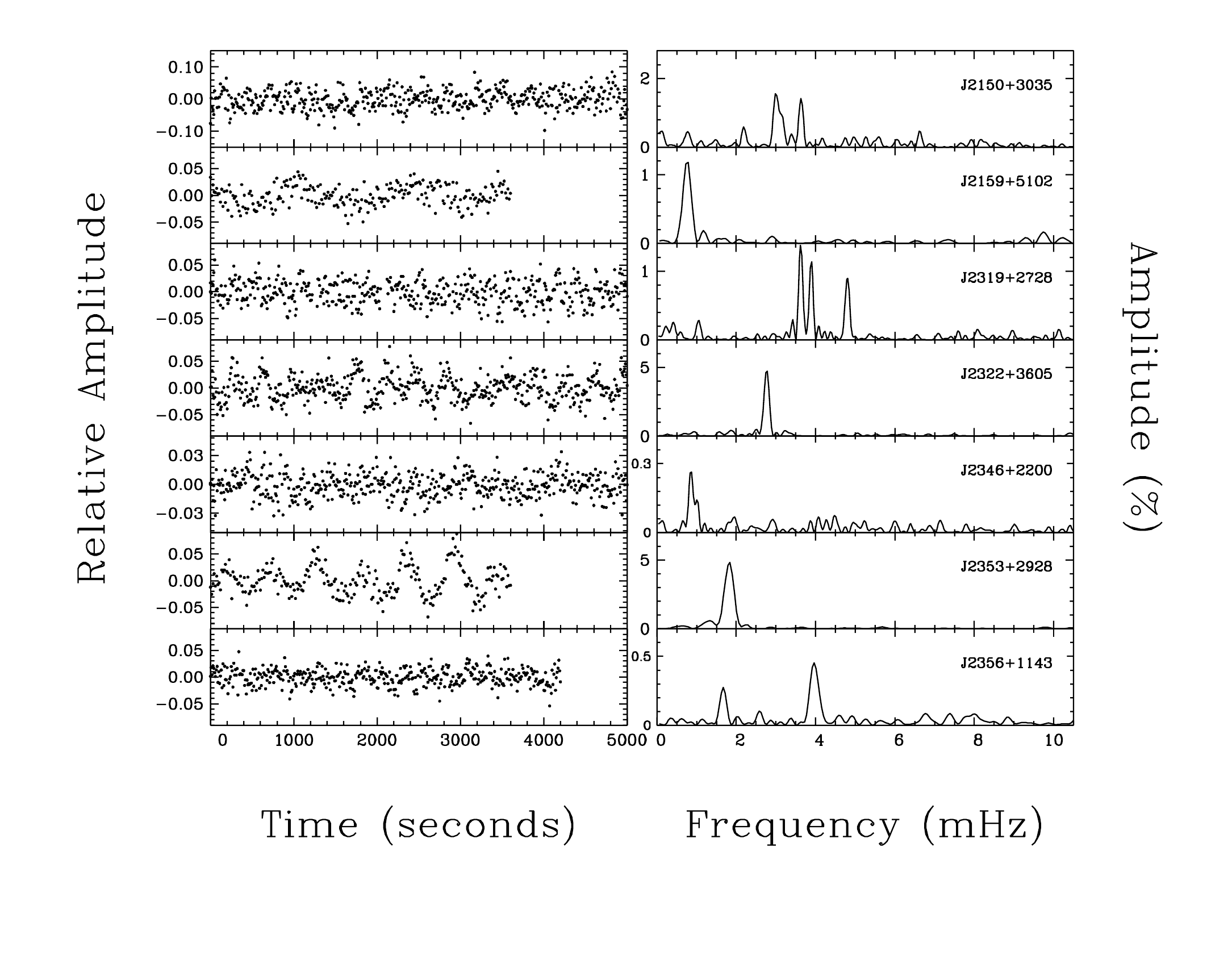}
 \caption{(Continued)}
\end{center}
\end{figure}
\setcounter{figure}{7}

\afterpage{\begin{longrotatetable}
\startlongtable
\begin{deluxetable}{lrcccrccccl}
\tablecaption{NOV Candidate Properties\label{tab:nov}}
\tabletypesize{\scriptsize}
\tablehead{
\colhead{WD} & \colhead{Gaia DR2 Source} & \colhead{R.A.} & \colhead{Dec.} & \colhead{Phot. Strip} & \colhead{$\Te$} & \colhead{$M$} & \colhead{$G$} & \colhead{$u$} & \colhead{Precision}  & \colhead{DA Classification}\\ 
\colhead{} & \colhead{} & \colhead{(J2015.5)} & \colhead{(J2015.5)} & \colhead{} & \colhead{(K)} & \colhead{(\msun)} & \colhead{} & \colhead{} & \colhead{(\%)} & \colhead{}
} 

\startdata
J0031$+$1239 & 2778812676229535616 & 00:31:51.29 & $+$12:39:45.04 & No & 12005 $\pm$ 91 & 0.577 $\pm$ 0.008 & 16.4 & 16.7 \tablenotemark{\scriptsize a} & 1.8 & - \\ 
J0036$+$4356 & 387724053774415104 & 00:36:20.14 & $+$43:56:55.76 & Yes & 11480 $\pm$ 54 & 0.591 $\pm$ 0.009 & 16.7 & -  & 1.5 & - \\ 
J0037$+$5118 & 415684119076509056 & 00:37:15.30 & $+$51:18:44.34 & No & 12421 $\pm$ 113 & 0.790 $\pm$ 0.011 & 17.0 & -  & 1.8 & - \\ 
J0056$+$4410 & 377231345590861824 & 00:56:56.67 & $+$44:10:29.62 & Yes & 11004 $\pm$ 66 & 0.655 $\pm$ 0.008 & 16.4 & 16.7 \tablenotemark{\scriptsize a} & 1.5 & - \\ 
J0056$+$4410 & 377231139432432384 & 00:56:57.17 & $+$44:10:18.55 & Yes & 11798 $\pm$ 75 & 0.568 $\pm$ 0.006 & 16.0 & 16.3 \tablenotemark{\scriptsize a} & 1.5 & - \\ 
J0135$+$5722 & 412839403319209600 & 01:35:17.69 & $+$57:22:47.67 & Yes & 12576 $\pm$ 89 & 1.156 $\pm$ 0.004 & 16.7 & 16.8 \tablenotemark{\scriptsize a} & 7.8 & \citet{kilic2020} \\ 
J0307$+$3157 & 135715232773818368 & 03:07:41.88 & $+$31:57:34.26 & Yes & 11560 $\pm$ 51 & 0.587 $\pm$ 0.009 & 16.4 & 16.8 \tablenotemark{\scriptsize b} & 1.8 & \citet{kawka2006} \\ 
J0341$-$0322 & 3249740657527506048 & 03:41:54.43 & $-$03:22:39.46 & Yes & 11804 $\pm$ 92 & 0.611 $\pm$ 0.006 & 15.3 & -  & 0.9 & \citet{Gianninas2011} \\ 
J0345$+$1940 & 63054590968017408 & 03:45:12.05 & $+$19:40:24.30 & No & 12367 $\pm$ 82 & 0.743 $\pm$ 0.004 & 14.2 & -  & 0.7 & - \\ 
J0408$+$2323 & 53716846734195328 & 04:08:03.02 & $+$23:23:42.48 & Yes & 12071 $\pm$ 110 & 1.038 $\pm$ 0.012 & 17.3 & -  & 5.6 & - \\ 
J0501$+$3323 & 184735992329821312 & 05:01:42.72 & $+$33:23:44.46 & No & 12107 $\pm$ 90 & 0.633 $\pm$ 0.008 & 16.1 & -  & 1.4 & - \\ 
J0533$+$6057 & 283096760659311744 & 05:33:45.33 & $+$60:57:50.14 & Yes & 11468 $\pm$ 61 & 0.585 $\pm$ 0.006 & 15.8 & 16.1 \tablenotemark{\scriptsize a} & 1.4 & \citet{kleinman2013} \\ 
J0538$+$3212 & 3447991090873280000 & 05:38:58.04 & $+$32:12:28.39 & Yes & 12457 $\pm$ 154 & 0.996 $\pm$ 0.012 & 17.5 & -  & 5.1 & - \\ 
J0626$+$3213 & 3439162768415866112 & 06:26:13.28 & $+$32:13:11.33 & Yes & 11660 $\pm$ 94 & 0.563 $\pm$ 0.007 & 16.2 & -  & 2.5 & - \\ 
J0634$+$3848 & 945007674022721280 & 06:34:16.58 & $+$38:48:55.09 & Yes & 12210 $\pm$ 106 & 0.926 $\pm$ 0.008 & 15.8 & -  & 2.3 & \citet{guo2015} \\ 
J0657$+$7341 & 1114813977776610944 & 06:57:11.11 & $+$73:41:44.62 & Yes & 12625 $\pm$ 118 & 1.142 $\pm$ 0.008 & 17.7 & -  & 3.9 & - \\ 
J0717$+$6214 & 1087442842689746048 & 07:17:07.39 & $+$62:14:07.53 & Yes & 11222 $\pm$ 67 & 0.648 $\pm$ 0.007 & 15.8 & -  & 3.8 & \citet{mickaelian2010} \\ 
J0739$+$2008 & 672816969200760064 & 07:39:19.79 & $+$20:08:29.53 & No & 12283 $\pm$ 145 & 0.710 $\pm$ 0.009 & 16.0 & 16.2 \tablenotemark{\scriptsize a} & 2.8 & - \\ 
J0748$-$0323 & 3080844435869554176 & 07:48:41.91 & $-$03:23:34.81 & Yes & 11391 $\pm$ 29 & 0.611 $\pm$ 0.004 & 15.7 & -  & 1.3 & - \\ 
J0751$+$1120 & 3150770626615542784 & 07:51:41.46 & $+$11:20:29.07 & Yes & 11728 $\pm$ 58 & 0.558 $\pm$ 0.007 & 16.4 & 16.8 \tablenotemark{\scriptsize a} & 2.6 & \citet{kilic2020} \\ 
J1157$+$5110 & 791138993175412480 & 11:57:22.38 & $+$51:10:13.11 & No & 12075 $\pm$ 121 & 0.597 $\pm$ 0.008 & 16.3 & 16.7 \tablenotemark{\scriptsize a} & 2.3 & - \\ 
J1243$+$4805 & 1543370904111505408 & 12:43:41.62 & $+$48:05:34.94 & Yes & 12716 $\pm$ 91 & 0.966 $\pm$ 0.006 & 17.0 & 17.2 \tablenotemark{\scriptsize a} & 4.0 & - \\ 
J1308$+$5754 & 1566530913957066240 & 13:08:48.48 & $+$57:54:37.03 & Yes & 11622 $\pm$ 83 & 0.704 $\pm$ 0.010 & 16.8 & 17.2 \tablenotemark{\scriptsize a} & 4.3 & \citet{kilic2020} \\ 
J1322$+$0757 & 3719371829283488768 & 13:22:47.58 & $+$07:57:29.60 & No & 12460 $\pm$ 67 & 0.702 $\pm$ 0.008 & 16.4 & 16.7 \tablenotemark{\scriptsize a} & 3.4 & - \\ 
J1557$-$0701 & 4349734833473621248 & 15:57:26.24 & $-$07:01:21.23 & Yes & 11792 $\pm$ 72 & 0.607 $\pm$ 0.006 & 16.1 & -  & 2.7 & - \\ 
J1559$+$2635 & 1316268323580640256 & 15:59:55.25 & $+$26:35:19.06 & No & 12266 $\pm$ 114 & 0.706 $\pm$ 0.007 & 16.3 & 16.6 \tablenotemark{\scriptsize a} & 2.6 & - \\ 
J1607$+$2933 & 1317275544951049472 & 16:07:24.37 & $+$29:33:23.51 & Yes & 10897 $\pm$ 38 & 0.685 $\pm$ 0.004 & 15.6 & 16.0 \tablenotemark{\scriptsize a} & 0.8 & \citet{stephenson1992} \\ 
J1617$+$1129 & 4454017257893306496 & 16:17:09.38 & $+$11:29:01.43 & Yes & 11696 $\pm$ 64 & 0.711 $\pm$ 0.007 & 16.5 & 16.8 \tablenotemark{\scriptsize a} & 0.9 & \citet{pauli2006} \\ 
J1626$+$2533 & 1304081783374935680 & 16:26:59.55 & $+$25:33:27.60 & No & 13313 $\pm$ 203 & 1.143 $\pm$ 0.007 & 17.6 & 17.7 \tablenotemark{\scriptsize a} & 5.8 & - \\ 
J1635$+$5053 & 1411867767238390912 & 16:35:05.49 & $+$50:53:59.78 & Yes & 10416 $\pm$ 41 & 0.554 $\pm$ 0.005 & 16.3 & 16.7 \tablenotemark{\scriptsize a} & 1.0 & \citet{kilic2020} \\ 
J1643$-$0953 & 4337833650892408448 & 16:43:15.16 & $-$09:53:05.43 & Yes & 11443 $\pm$ 81 & 0.477 $\pm$ 0.008 & 16.5 & -  & 1.7 & - \\ 
J1643$+$6328 & 1631796309274519040 & 16:43:50.51 & $+$63:28:29.16 & Yes & 12380 $\pm$ 121 & 0.838 $\pm$ 0.008 & 17.0 & 17.2 \tablenotemark{\scriptsize a} & 1.2 & \citet{kilic2020} \\ 
J1643$+$1118 & 4447022061837071744 & 16:43:54.06 & $+$11:18:49.28 & No & 12302 $\pm$ 156 & 0.646 $\pm$ 0.009 & 16.5 & 16.7 \tablenotemark{\scriptsize a} & 2.4 & - \\ 
J1652$+$4110 & 1353302001211658368 & 16:52:00.69 & $+$41:10:31.36 & Yes & 11124 $\pm$ 76 & 0.894 $\pm$ 0.008 & 17.1 & 17.4 \tablenotemark{\scriptsize a} & 1.2 & - \\ 
J1702$+$3905 & 1353355434900703616 & 17:02:41.82 & $+$39:05:58.25 & Yes & 10547 $\pm$ 44 & 0.681 $\pm$ 0.005 & 16.3 & 16.6 \tablenotemark{\scriptsize a} & 0.9 & \citet{kilic2020} \\ 
J1706$-$0837 & 4336571785203401472 & 17:06:18.45 & $-$08:37:52.44 & Yes & 12143 $\pm$ 249 & 1.163 $\pm$ 0.011 & 17.4 & -  & 4.2 & - \\ 
J1728$+$2053 & 4555079659441944960 & 17:28:45.69 & $+$20:53:40.98 & No & 12017 $\pm$ 38 & 0.621 $\pm$ 0.007 & 16.7 & 17.0 \tablenotemark{\scriptsize b} & 1.0 & - \\ 
J1805$+$1536 & 4498531123585093120 & 18:05:43.90 & $+$15:36:40.03 & Yes & 11342 $\pm$ 85 & 0.581 $\pm$ 0.010 & 16.7 & -  & 1.2 & - \\ 
J1813$+$4427 & 2114985726416563072 & 18:13:01.14 & $+$44:27:19.05 & Yes & 11149 $\pm$ 90 & 1.098 $\pm$ 0.007 & 17.7 & -  & 1.4 & - \\ 
J1854$+$0411 & 4281190419601308672 & 18:54:50.41 & $+$04:11:26.21 & Yes & 12472 $\pm$ 170 & 0.904 $\pm$ 0.012 & 17.3 & -  & 4.4 & - \\ 
J1857$+$3353 & 2092086476924423808 & 18:57:57.29 & $+$33:53:03.88 & Yes & 10477 $\pm$ 62 & 0.632 $\pm$ 0.009 & 16.8 & -  & 1.2 & - \\ 
J1910$+$7334 & 2265100885021724032 & 19:10:43.38 & $+$73:34:39.06 & No & 13119 $\pm$ 214 & 1.087 $\pm$ 0.008 & 17.7 & -  & 2.4 & - \\ 
J1928$+$1526 & 4321498378443922816 & 19:28:14.56 & $+$15:26:38.51 & Yes & 11543 $\pm$ 136 & 1.002 $\pm$ 0.013 & 17.8 & 18.0 \tablenotemark{\scriptsize a} & 6.4 & \citet{kilic2020} \\ 
J1949$+$4734 & 2086392484163910656 & 19:49:14.55 & $+$47:34:45.72 & No & 11911 $\pm$ 88 & 0.570 $\pm$ 0.006 & 16.1 & -  & 0.7 & - \\ 
J1950$+$7155 & 2263690864438162944 & 19:50:45.89 & $+$71:55:40.93 & Yes & 11451 $\pm$ 90 & 0.711 $\pm$ 0.008 & 16.7 & -  & 1.1 & \citet{voss2007} \\ 
J1954$+$0848 & 4298401105174809984 & 19:54:49.52 & $+$08:48:50.54 & Yes & 10594 $\pm$ 77 & 0.640 $\pm$ 0.012 & 16.9 & -  & 1.3 & - \\ 
J2001$+$2620 & 1835056216381670272 & 20:01:17.81 & $+$26:20:21.33 & Yes & 11643 $\pm$ 122 & 0.668 $\pm$ 0.010 & 16.9 & -  & 2.7 & - \\ 
J2014$+$8018 & 2292229788249205760 & 20:14:34.37 & $+$80:18:42.53 & Yes & 10591 $\pm$ 62 & 0.668 $\pm$ 0.007 & 16.5 & -  & 1.0 & - \\ 
J2017$+$4653 & 2083300584444566016 & 20:17:53.54 & $+$46:53:14.89 & No & 12141 $\pm$ 108 & 0.600 $\pm$ 0.008 & 16.6 & -  & 1.6 & - \\ 
J2030$+$1857 & 1815614965310875520 & 20:30:08.62 & $+$18:57:34.75 & Yes & 10511 $\pm$ 61 & 0.648 $\pm$ 0.009 & 16.7 & -  & 1.3 & - \\ 
J2032$+$4801 & 2083661675243196544 & 20:32:28.75 & $+$48:01:46.28 & Yes & 11939 $\pm$ 74 & 0.779 $\pm$ 0.006 & 16.6 & -  & 0.9 & - \\ 
J2045$+$3844 & 2063435712171048704 & 20:45:28.02 & $+$38:44:26.65 & Yes & 10629 $\pm$ 43 & 0.649 $\pm$ 0.006 & 15.8 & -  & 2.2 & - \\ 
J2049$+$4500 & 2163226700308494080 & 20:49:02.69 & $+$45:00:36.26 & Yes & 10993 $\pm$ 73 & 0.659 $\pm$ 0.007 & 15.6 & 15.9 \tablenotemark{\scriptsize a} & 0.6 & \citet{kilic2020} \\ 
J2053$+$2705 & 1845487489350432128 & 20:53:51.74 & $+$27:05:53.58 & Yes & 11178 $\pm$ 202 & 1.205 $\pm$ 0.011 & 18.3 & -  & 2.0 & - \\ 
J2054$+$2427 & 1842670231320998016 & 20:54:46.68 & $+$24:27:29.23 & No & 12534 $\pm$ 95 & 0.713 $\pm$ 0.006 & 15.9 & 16.1 \tablenotemark{\scriptsize a} & 1.0 & - \\ 
J2119$+$4206 & 1968901145520461568 & 21:19:01.61 & $+$42:06:16.46 & Yes & 11009 $\pm$ 102 & 0.450 $\pm$ 0.009 & 16.5 & -  & 0.9 & - \\ 
J2122$+$6600 & 2220815923910913920 & 21:22:31.89 & $+$66:00:42.62 & Yes & 10613 $\pm$ 55 & 0.708 $\pm$ 0.006 & 15.9 & -  & 0.7 & - \\ 
J2150$+$2205 & 1793328410074430464 & 21:50:07.49 & $+$22:05:56.32 & No & 12761 $\pm$ 113 & 0.856 $\pm$ 0.010 & 17.0 & 17.2 \tablenotemark{\scriptsize a} & 1.2 & - \\ 
J2305$+$5125 & 1995097319287822080 & 23:05:31.71 & $+$51:25:20.49 & Yes & 11458 $\pm$ 52 & 0.608 $\pm$ 0.005 & 15.7 & -  & 2.3 & - \\ 
J2312$+$4206 & 1930609656643838080 & 23:12:42.51 & $+$42:06:00.42 & Yes & 10445 $\pm$ 62 & 0.574 $\pm$ 0.009 & 16.8 & -  & 1.1 & - \\ 
J2318$+$1236 & 2811321837744375936 & 23:18:45.10 & $+$12:36:02.77 & Yes & 11710 $\pm$ 67 & 0.866 $\pm$ 0.005 & 15.4 & 15.7 \tablenotemark{\scriptsize a} & 0.6 & \citet{ferrario2015} \\ 
J2336$+$0335 & 2647884790098989568 & 23:36:17.00 & $+$03:35:08.12 & Yes & 11218 $\pm$ 83 & 0.549 $\pm$ 0.010 & 16.5 & 16.9 \tablenotemark{\scriptsize a} & 3.2 & - \\ 
J2341$+$5750 & 1998740551069600128 & 23:41:07.61 & $+$57:50:53.83 & No & 11793 $\pm$ 64 & 0.523 $\pm$ 0.005 & 15.8 & -  & 1.0 & - \\ 
J2347$+$5312 & 1993426577008368640 & 23:47:09.28 & $+$53:12:17.32 & Yes & 10639 $\pm$ 89 & 0.741 $\pm$ 0.011 & 17.0 & -  & 1.2 & - \\ 
J2356$+$0803 & 2746936704565640064 & 23:56:06.84 & $+$08:03:22.28 & No & 12030 $\pm$ 88 & 0.613 $\pm$ 0.008 & 16.1 & 16.4 \tablenotemark{\scriptsize a} & 1.1 & - \\ 
\enddata
\tablenotetext{a}{SDSS photometry.}
\tablenotetext{b}{CFIS photometry.}
\end{deluxetable}
\end{longrotatetable}} 

NOV targets in our sample, as well as the 18 additional objects above
the blue edge, are listed in Table \ref{tab:nov} with the same
information as before, in addition to the photometric precision limit
of each light curve and literature identifying the object as a DA, if
available. The precision limit corresponds to the light curve standard
deviation, and is a good indicator of the smallest detectable
amplitude. Also included in Table \ref{tab:nov} is a column indicating
whether or not the object is located within the photometric
instability strip, to help distinguish objects from our prior
selection based on the spectroscopic instability strip.

\subsection{The Empirical ZZ Ceti Instability Strip}\label{sec:strip}

\begin{figure}
\begin{center}
 \includegraphics[angle=270,width=0.82\columnwidth,clip,trim=0.95in 0.5in 1in 0.66in]{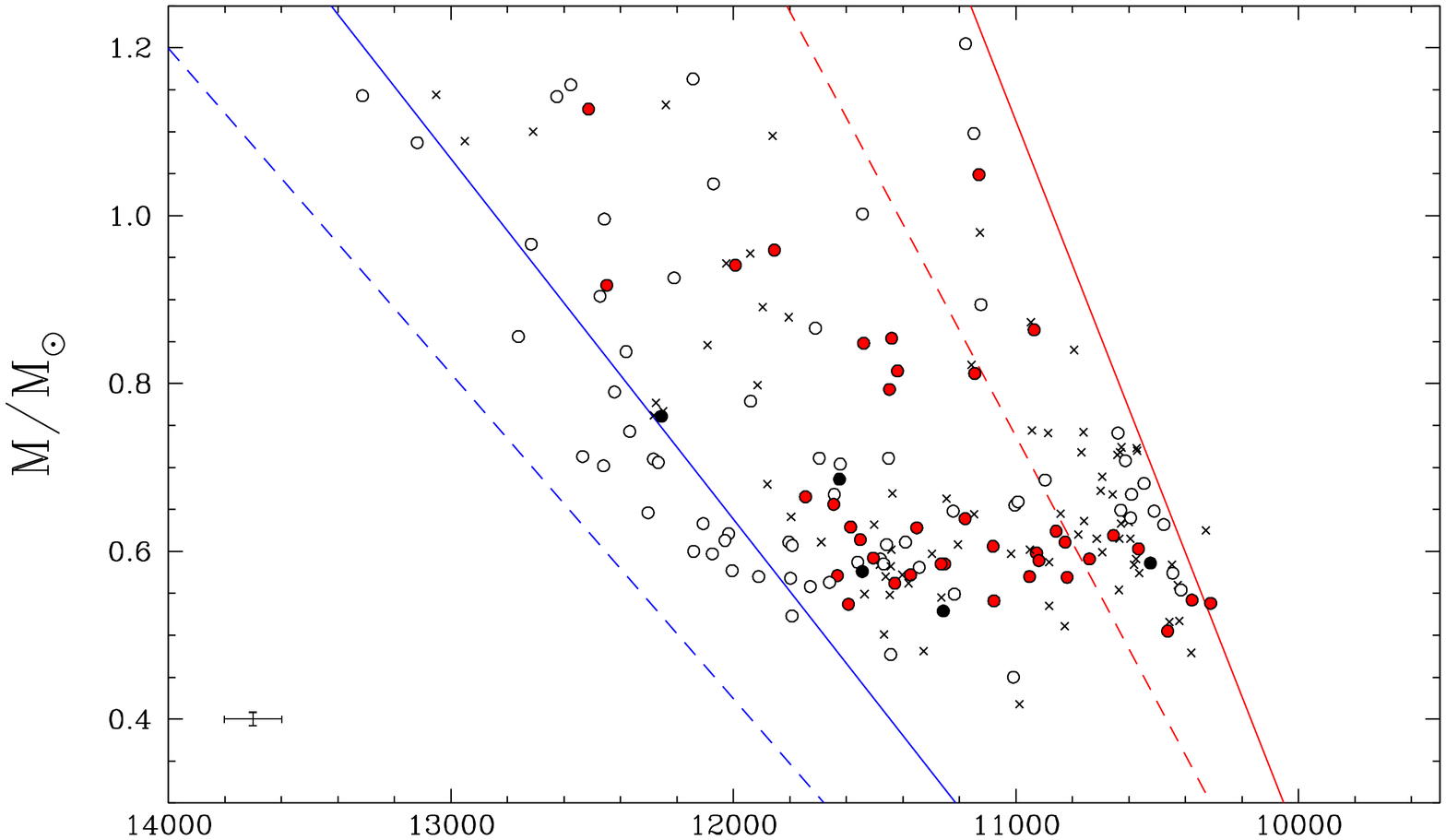}
 \includegraphics[angle=270,width=0.82\columnwidth,clip,trim=1.1in 0.5in 0.45in 0.66in]{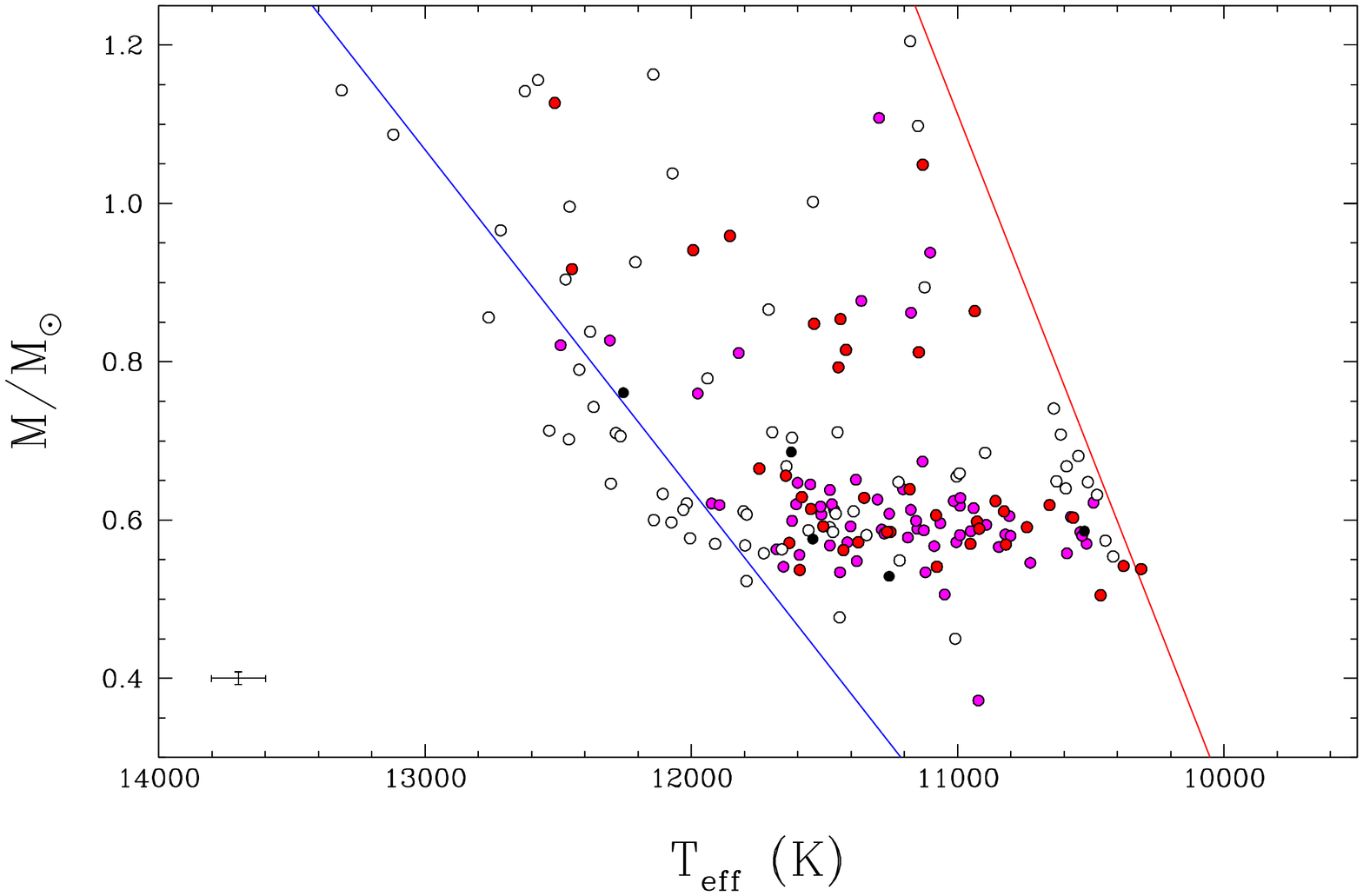} 
 \caption{Top: $M - \Te$ distribution for the 172 ZZ Ceti candidates
   and 18 objects previously selected for high-speed photometric
   follow-up. Different symbols are used to indicate new ZZ Ceti stars
   (red circles), possible pulsators (black circles), NOV objects
   (white circles), and remaining candidates yet to be observed (cross
   symbols). The empirical spectroscopic (dashed lines) and
   photometric (solid lines) ZZ Ceti instability strips taken from
   Figure \ref{fig:fullsamp} are also reproduced. Bottom: Same as top
   panel, but with the addition of the previously known ZZ Ceti stars
   within 100 pc from the Sun (magenta circles); for clarity, only the
   photometric instability strip and observed candidates are shown.}
   \label{fig:survey}
\end{center}
\end{figure}

The $M - \Te$ distribution for the 173 ZZ Ceti candidates and the 18
objects previously selected for high-speed photometric follow-up is
shown in the top panel of Figure \ref{fig:survey}, along with the
spectroscopic and photometric instability strips discussed in Section
\ref{sec:sample}. The new ZZ Ceti stars, possible pulsators, NOV
objects, and remaining candidates yet to be observed are identified
with different symbols in the figure. A first obvious result is the
presence of a large number of NOV white dwarfs within the ZZ Ceti
instability strip, suggesting that the strip is not pure. We postpone
our discussion of these objects to Section \ref{sec:nov}.

In the bottom panel of Figure \ref{fig:survey}, we show the same
distribution of objects in the $M - \Te$ diagram, but this time by
also including the previously known ZZ Ceti pulsators within 100 pc
from the Sun, already displayed in the bottom panel of Figure
\ref{fig:fullsamp}. To get a clearer picture, we removed from this
figure the location of the empirical spectroscopic instability
strip. So far, all of our new ZZ Ceti stars are found well within the
previously-defined empirical photometric instability strip, with the
bulk of them located near the average mass of white dwarfs around
$\sim$0.6 \msun. More interestingly, we have identified 11 new massive
($M\gtrsim 0.75$ \msun) pulsators, bringing a noticeable contribution
to the 7 currently known massive ZZ Ceti stars \citep{corsico2019}
contained within the volume of our sample. The relatively small number
of previously known massive pulsators can be attributed to a
well-known observational bias. Indeed, ZZ Ceti stars have been
previously identified mostly from magnitude-limited surveys. In such
surveys, massive white dwarfs are usually underrepresented due to
their intrinsic smaller radii and lower luminosities compared to their
normal mass counterparts \citep{giammichele2012}. In contrast, our
volume-limited survey provides instead an unbiased sample where
completeness issues are better controlled.

For similar reasons, less massive white dwarfs, with their larger
radii and higher luminosities, will be sampled at much larger
distances in magnitude-limited surveys, and will thus be
overrepresented. This can be appreciated by comparing the number of
low-mass ($M\lesssim0.4$ \msun) white dwarfs in Figure
\ref{fig:survey} with the number observed in Figure 11 of
\citet{bergeron2019}, which is based on the white dwarfs contained in
the MWDD, most of which have been discovered in magnitude-limited
surveys. Hence, not surprisingly, our survey has revealed no
additional low-mass pulsators. The only previously known low-mass ZZ
Ceti star in Figure \ref{fig:survey} is HS 1824+6000
\citep{steinfadt2008}, whose spectroscopic mass (3D-corrected) is also
low, $M\sim0.45$ \msun\ according to \citet{Gianninas2011}.

Also worth mentioning is our discovery of two new ultra-massive ($M
\gtrsim 1.0$ \msun) pulsators, J0551$+$4135 ($1.127\pm0.005$ \msun)
and J0204$+$8713 ($1.049\pm0.0015$ \msun). At the time of writing this
paper, only three other ultra-massive ZZ Ceti stars have been
confirmed: BPM 37093 with $M \sim 1.1$ \msun\ \citep{kanaan1992}, SDSS
J084021.23+522217.4 with $M \sim 1.16$ \msun\ \citep{curd2017}, and GD
518 with $M \sim 1.24$ \msun\ \citep{hermes2013}. Our new massive
and ultra-massive pulsators represent objects of interest for
asteroseismologic studies of the process of core crystallization within
the instability strip \citep{romero2013}. J0551$+$4135 is of particular
interest since ultra-massive ZZ Ceti stars with $M \gtrsim 1.1$ \msun~are 
expected to have a very large portion of their mass in the crystallized
phase \citep{degeronimo2019}, and 2-minute-cadence observations from TESS 
\citep[{\it Transiting Exoplanet Survey Satellite},][]{ricker2015} 
are available for this object.

We end this section with a few words regarding the exact location of
the empirical ZZ Ceti instability strip based on our photometric
survey. The bottom panel of Figure \ref{fig:survey} shows all variable
stars, both new and known, to be within the photometric instability
strip previously defined in Figure \ref{fig:fullsamp}, within the
uncertainties. Moreover, new pulsators found near the red edge of the
strip show diminishing amplitudes as they approach the edge itself
(further discussed in Section \ref{sec:puls}), strengthening our
assumption of its location. By the same token, the 18 NOV objects
observed above the blue edge are particularly useful to pinpoint its
exact location. Given the results shown here, we do not feel it is
necessary to revise the location of the blue edge of the photometric
instability strip. This in turn suggests an excellent internal
consistency between the spectroscopic and photometric determinations,
with the understanding that one is shifted in temperature with respect
to the other.

\subsection{Non-variability and the Purity of the ZZ Ceti Instability Strip}\label{sec:nov}

In this section, we discuss the purity of the ZZ Ceti instability
strip with respect to our findings, summarized in the top panel of
Figure \ref{fig:survey}. There are several aspects to consider when
assessing the purity of the instability strip, the most important of
which are the precision limits of the high-speed photometric
observations, and the accuracy and precision of the physical parameter
measurements\footnote{Statistically speaking, the precision of the
  method describes random errors, a measure of statistical
  variability, repeatability, or reproducibility of the measurement,
  while the accuracy represents the proximity of the measurements to
  the true value being measured.}. In our case, we also have to
consider the atmospheric composition of the candidates.

We find in our survey 47 NOV white dwarfs within the photometric
instability strip, 9 of which have a DA spectral type published in the
literature, while 8 more have recently been confirmed to be DAs by
\citet{kilic2020}. We note, however, that two of the published DA
spectral classifications are dubious. J0717$+$6214 (GD 449) was
classified as ``DA:'' by \citet{mickaelian2010}, where the colon
implies an uncertain spectral type. It would be difficult to
misclassify such a bright ($G\sim 15.8$) DA star in the temperature
range where ZZ Ceti stars are found, given that the Balmer lines reach
their maximum strength around $\Te\sim13,000$~K. We suspect the
authors may have detected an H$\alpha$ absorption feature in a
helium-rich DBA white dwarf. The second object, J1950$+$7155 (HS
1951$+$7147), is classified as DA in Simbad, with a reference to
\citet{voss2007}, who reported in their Table 1 this object to be a
non-variable white dwarf. The atmospheric parameters for this object
were derived from BUSCA photometry using pure hydrogen models (see
Voss et al.~for details), although we find no evidence for a firm DA
spectral classification in their analysis. In fact, there was no
follow-up on HS 1951$+$7147 in the spectroscopic analysis of DA white
dwarfs from the ESO SN Ia Progenitor Survey published by
\citet{koester2009}. As there is no spectroscopic evidence for the DA
classification for this object, the possibility of a helium-atmosphere
remains.

We also realized after the fact that J2318$+$1236 (KUV 23162+1220) is
a highly magnetic ($B_p\sim 45$~MG) DA white dwarf \citep[][see the
  spectrum in Figure 5 of \citealt{Gianninas2011}]{ferrario2015}. It
has been suggested that the presence of a strong magnetic field might
have a dramatic effect on the driving mechanism of the pulsations
\citep[see Section 3.4 of][]{tremblay2015}. This suggestion has been
reinforced by \citet{Gentile2018} who reported the convincing case of
a $\Te\sim10,000$~K DA white dwarf (WD~2105$-$820, L24-52) in which
atmospheric convection has been suppressed by the presence of even a
weak magnetic field \citep[$B_p\sim 56$~kG,][]{Landstreet2012}. Since
the driving mechanism in ZZ Ceti stars is located at the bottom of the
hydrogen convective zone, it is reasonable to assume that magnetic DA
stars should not pulsate. Our photometric observations of J2318$+$1236
certainly support this interpretation. It is thus possible that
additional NOV objects in our sample are magnetic DA white dwarfs,
even weakly magnetic.

Among the 47 NOV white dwarfs within the instability strip, 18 are
confirmed to be hydrogen-rich through their $u$-band photometry (9 of
these 18 also have a firm DA spectral type, including the magnetic
DA). Excluding the genuine DA stars discussed above, this leaves 26
NOV white dwarfs within the strip that could possibly have a helium
atmosphere or be magnetic; these can only be confirmed with additional
spectroscopic or $u$-band photometric observations. We thus end up
with 20 NOV white dwarfs within the instability strip that are either
hydrogen-rich through their $u$-band photometry or that are classified
as genuine DA stars, excluding the magnetic white dwarf. These are the
offending NOV objects we need to explain. In every case, there is
always the remote possibility for pulsations in a ZZ Ceti star to be
hidden from us due to geometric considerations (see, for example, HS
1612$+$5528 discussed in \citealt{Gianninas2011}).

While \citet{bergeron2004} argued that the ZZ Ceti instability strip
is pure --- i.e.~devoid of non-variable white dwarfs --- when analyzed
using the spectroscopic technique, our study is the first assessment
of its purity based on the detailed photometric approach. \citet[][see
  also \citealt{tremblay2019} and \citealt{gentile2019}]{genest2019}
discussed at length the accuracy and precision of both the
spectroscopic and photometric techniques. They argued that even though
the photometric approach yields physical parameters that are more
accurate, the spectroscopic method is probably more precise. For
instance, while differences in spectroscopic and photometric
temperatures in Figure \ref{fig:comp_VINCENT} are of the order of 5\%
or less, on average, there are cases where these differences can reach
15\% or more.

We can explore these discrepancies more quantitatively by comparing
our photometric parameters with those obtained from spectroscopy for
some of the offending NOV objects within the instability strip with
optical spectra available to us. For instance, for J0341$-$0322
\citep[LP 653-26; spectrum from][]{Gianninas2011}, we obtain a
spectroscopic temperature of $T_{\rm spec}=12,807$~K using our
ML2/$\alpha=0.7$ models, a value 8.5\% higher than our photometric
temperature given in Table \ref{tab:nov} ($T_{\rm
  phot}=11,804$~K). With a (3D-corrected) spectroscopic mass of 0.64
\msun, this white dwarf is thus located above the empirical
spectroscopic instability strip. Similarly, we find that J0533+6057
\citep[SDSS J053345.32+605750.3; spectrum from][]{kleinman2013} and
J1617+1129 \citep[HS 1614+1136; spectrum from][]{koester2009} have
spectroscopic temperatures of $T_{\rm spec}=13,130$~K (with $T_{\rm
  phot}=11,468$~K) and $T_{\rm spec}=13,970$~K (with $T_{\rm
  phot}=11,696$~K), respectively, both significantly above the
spectroscopic instability strip. An even more extreme case is that of
J1243$+$4805 \citep[HS 1241+4821; SDSS spectrum from][]{kleinman2013},
for which we obtain $T_{\rm spec}=14,838$~K, a value more than 2000 K
hotter than our photometric temperature of $T_{\rm
  phot}=12,716$~K. Finally, \citet{kawka2006} report a spectroscopic
temperature of $T_{\rm spec}=13,300$~K for J0307$+$3157 (NLTT 9933),
while we obtain $T_{\rm phot}=11,560$~K. Hence, most of the
spectroscopic temperatures push these NOV objects above the blue edge
of the spectroscopic instability strip, suggesting that the
photometric temperatures might sometimes be underestimated.

It is worth noting in this context that among the new ZZ Ceti stars
listed in Table \ref{tab:var}, 19 are known to be DA white dwarfs.
\citet{limoges2015} obtained spectroscopic parameters ($\Te$ and $M$)
for 3 of those DA stars (J10042+2438, J19033+6035, and J18435+2740)
that place them well within the ZZ Ceti instability strip. We also
have spectra for 13 DA stars from the analysis of \citet{kilic2020},
and even though most of these are classification spectra with low
signal-to-noise ratios, the spectroscopic parameters obtained from the
best quality spectra also place them within the strip.  This
reinforces the idea that the NOV objects discussed above represent
cases where the photometric parameters suffer from large errors.

Also, we cannot exclude that in some cases, the differences between
spectroscopic and photometric temperatures may be explained in terms
of unresolved double degenerate binaries. Indeed, \citet{workshop2018}
showed that the most extreme differences in physical parameters ($\Te$
and $M$) tend to be associated with double DA white dwarf binaries,
for which the measured radii inferred from the photometric technique
are overestimated --- and thus the masses are underestimated --- due
to the presence of two stars, while the spectroscopic masses remain
relatively unaffected. J2119$+$4206, our lowest-mass NOV candidate,
seems to be such a case. It is also possible to have an unresolved
double DA+DC binary, where the DC star dilutes the hydrogen lines of
the DA component of the system, making the object appear as a massive
DA white dwarf when analyzed with the spectroscopic technique. An
excellent example is the DA star G122-31 --- also discussed by
\citet{workshop2018} --- which \citet{harris2013} reported as being an
unresolved degenerate binary. The spectroscopic parameters for this
object are $\Te=28,080$~K and $\log g=8.97$ (or $M=1.19$ \msun), while
the photometric values are significantly different, $\Te=14,648$~K and
$\log g=8.53$ (or $M=0.95$ \msun).

The bottom line of the above discussion is that we need a combined
spectroscopic and photometric investigation of our NOV candidates for
any serious discussion of the purity of the ZZ Ceti instability
strip. Therefore we cannot conclude at this stage that the strip
contains a significant number of non-variable DA white dwarfs.

Finally, we look at the confirmed ZZ Ceti stars to estimate the
likeliness of pulsations being hidden within photometric noise for the
NOV candidates. As discussed in the next section, pulsators located
very close to the edges of the instability strip typically show the
smallest amplitudes, sometimes as small as 0.1\%. Given that our
typical photometric precision is about 3.4\% for the average {\it
  Gaia} magnitude $\langle G \rangle = 16.5$ of our sample, detecting
such small pulsations in fainter objects is unlikely with our
observational capabilities. As we move further away from the edges and
toward the center of the strip, ZZ Ceti stars tend to have larger
amplitudes, and the likelihood of pulsations being smaller than our
photometric precision limit decreases. Another possibility is to have
observed the candidate amid a beat caused by two or more oscillation
  modes interfering destructively with each other. For example, in
the case of J0324$+$6020 (see the light curve in Figure
\ref{fig:lcfft}), we observed a beat lasting well over an hour, during
which the pulsation amplitude drops to a nearly undetectable
level. For fainter candidates, this could easily translate into
observing no pulsations. Ultimately, our NOV candidates will have to
be reobserved with better precision to better constrain their
  non-variability.  Longer light curves would also be more sensitive
  to small-amplitude pulsations. In particular, candidates located
near the edges will require higher performance facilities, or longer
observations, than what is offered at the Mont-M\'egantic Observatory.

\subsection{Pulsational Properties}\label{sec:puls}

ZZ Ceti white dwarfs exhibit a wide variety of light curves, and the
investigation of their periods, amplitudes, and nonlinearities can
reveal a wealth of information in the context of asteroseismological
studies. Of particular interest in this section is how these
characteristics evolve empirically across the ZZ Ceti instability
strip. Many global patterns have been established some time ago
\citep{robinson1979,fontaine1982}, such as the inverse correlation
between effective temperature and period \citep{winget1982}. A
temperature-amplitude relationship was also discussed by
\citet{kanaan2002} and \citet{mukadam2006}. In particular, Mukadam et
al.~have shown that the amplitudes increase with decreasing $\Te$,
reaching a maximum near the cooler half of the strip, after which the
amplitudes start to drop toward the red edge. While these
temperature-dependent relations have proven to hold true, it has been
demonstrated since then that they also depend on surface gravity
\citep{fontaine2008}. More recently, \citet{hermes2017} analyzed
a sample of 27 ZZ Ceti stars using the space-based observations
taken by the \textit{Kepler} telescope. The extended duration of the
light curves allowed to confirm what appears to be a new phase in the 
evolution of DAVs as they cool past the center of the instability strip:
aperiodic outbursts increasing the mean stellar flux by a few to 15\%, 
over several hours, and recur sporadically on a timescale of days.
Here, we take a fresh look at the ZZ Ceti ensemble characteristics using 
our sample of new pulsators.

\begin{figure}
\begin{center}
 \includegraphics[width=1\columnwidth,clip,trim=1.2in 0 1.5in 1.5in]{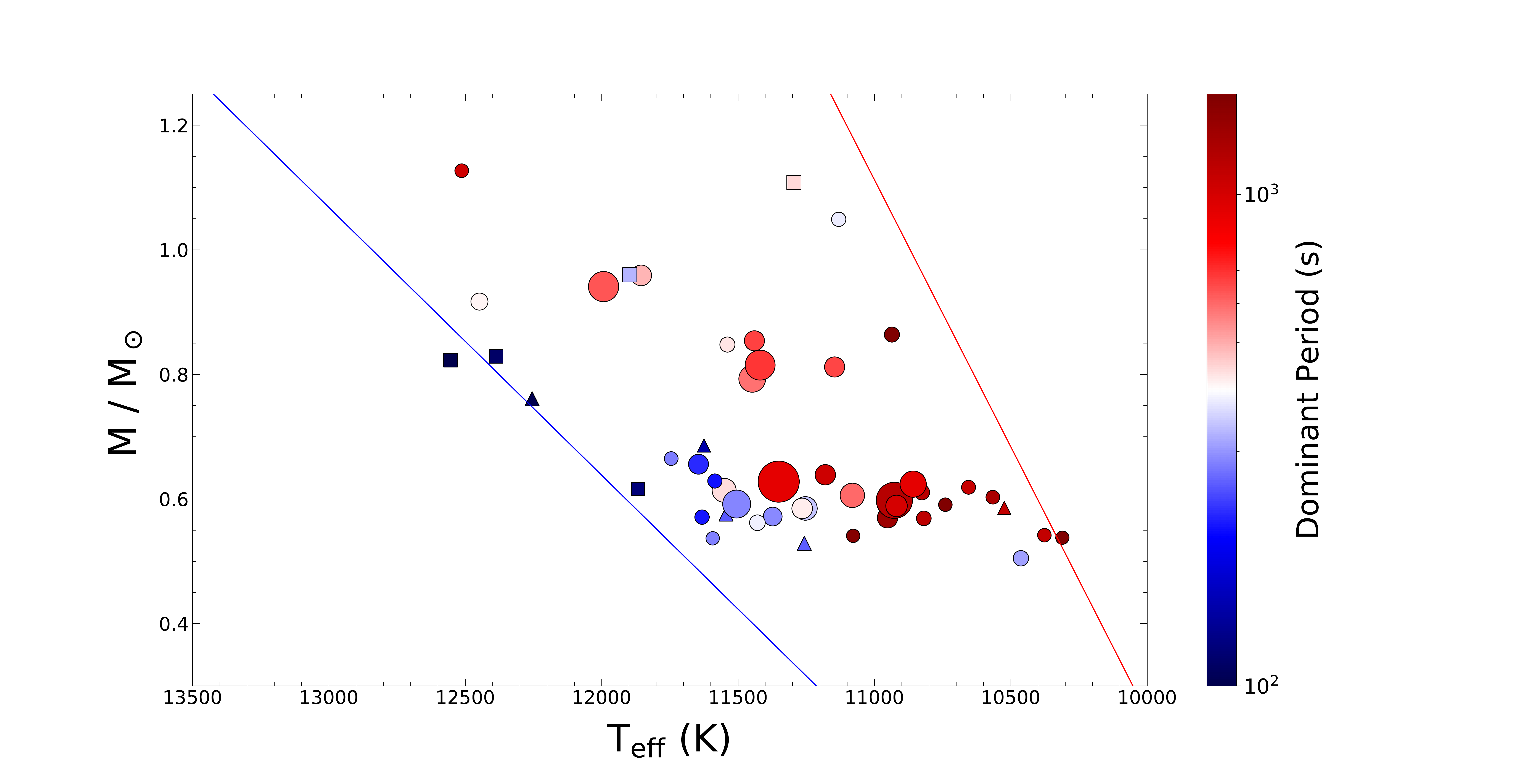}
\caption{Logarithmic color map of the dominant periods in the $M -
  \Te$ plane for our new ZZ Ceti white dwarfs (circles) and possible
  pulsators (triangles). Also displayed are a few previously known ZZ
  Ceti stars (squares) discussed in the text. The size of every object
  gives a measure of the amplitude of their dominant period, linearly
  scaling from 0.05 to 30\%.}
\label{fig:periods}
\end{center}
\end{figure}

Figure \ref{fig:periods} shows a color map in the $M - \Te$ plane for
the dominant periods ($P_d$) present in the light curves of our new
and possible pulsators, where the size of every symbol scales
according to the amplitude of the object's dominant period. Also
included in the figure are a few previously known ZZ Ceti stars of
interest, which will be discussed below. We detect periods between 195
and 1600~s, with a clear evolution from small periods near the blue
edge of the strip to longer values as we approach the red edge, with a
few exceptions: the two ultra-massive pulsators and two other objects
near the red edge. These will be discussed further below. Massive
($M\gtrsim0.75$~\msun) pulsators also follow the general trend,
although most of their periods are found within a narrow range from
350 to 600~s. In the first studies of the ensemble characteristics of
massive ZZ Ceti white dwarfs, \citet{castanheira2013} suggested a mode
selection mechanism preventing periods around 500~s due to a lack of
observed pulsations near this value (see their Figure 5). While mode
trapping is predicted to be more important for massive pulsators
\citep{brassard1992}, our results go against the idea of a particular
phenomenon completely suppressing periods between 400 to 600~s.

The odd pulsator J2319$+$2728, located close to the cool edge of the
ZZ Ceti instability strip at $M\sim0.5$~\msun, seemingly stands out
from the general trend of increasing periods, with $P_d=277$~s. 
A similar object (SDSS J2350$-$0054) was reported by \citet{mukadam2004},
who found no obvious explanation for its peculiar properties. In the 
case of J2319$+$2728, the Pan-STARRS photometry was found to possibly 
be contaminated by a neighboring luminous star, which most likely 
results in an overestimation of the stellar radius --- and thus 
an underestimation of the stellar mass --- when using the photometric
technique. The impact on $\Te$ is presumably less important, given
that this ZZ Ceti star is still located within the boundaries of the
instability strip, but the temperature remains affected
nonetheless. Given its pulsational properties, we suspect the object
actually lies among the bulk of our new pulsators, closer to the blue
edge.

Another noteworthy case is the ultra-massive pulsators. J0551$+$4135
shows a period ($P_d=809$~s) much longer than the periods found in
other massive ZZ Ceti stars in the same temperature range, and
J0204$+$8713 shows the exact opposite with a much shorter period
($P_d=330$~s) than found in cool massive pulsators. As an attempt to
discern a trend among the ultra-massive pulsators in the $M - \Te$
plane, we included in the color map of Figure \ref{fig:periods} two of
the three aforementioned ultra-massive objects, using our own
photometric measurements of their Pan-STARRS photometry (see Section
\ref{sec:strip}). With GD 518 at $\Te=11,295$~K and
$M=1.108$~\msun\ \citep[$P_d \sim 442$~s,][]{hermes2013}, SDSS
J084021.23$+$522217.4 at $\Te=11,897$~K and
$M=0.962$~\msun\footnote{We note here that the photometric mass for
  this object is below 1 \msun. We have optical spectra for 2 of our new
  ZZ Ceti stars (from \citealt{kilic2020}) with $M_{\rm
    phot}>0.9$~\msun, and although the spectrum of J0856$+$6206 is too
  noisy for a proper spectroscopic analysis, we obtain for J1812$+$4321
  a spectroscopic mass of $M_{\rm spec}=0.99$~\msun\ (compared to
  $M_{\rm phot}=0.917$~\msun), possibly adding another ultra-massive
  white dwarfs to the sample.} \citep[$P_d \sim 326$~s,][]{curd2017},
the four ultra-massive ZZ Ceti stars appear to show diminishing
periods as they cool down the strip. A more detailed study of these
objects will be required to confirm this phenomenon, as it would go
against the general trend observed in all other ZZ Ceti white dwarfs.

\begin{figure}
\begin{center}
 \includegraphics[angle=270,width=0.82\columnwidth,clip,trim=0.95in 0.5in 0.25in 0.66in]{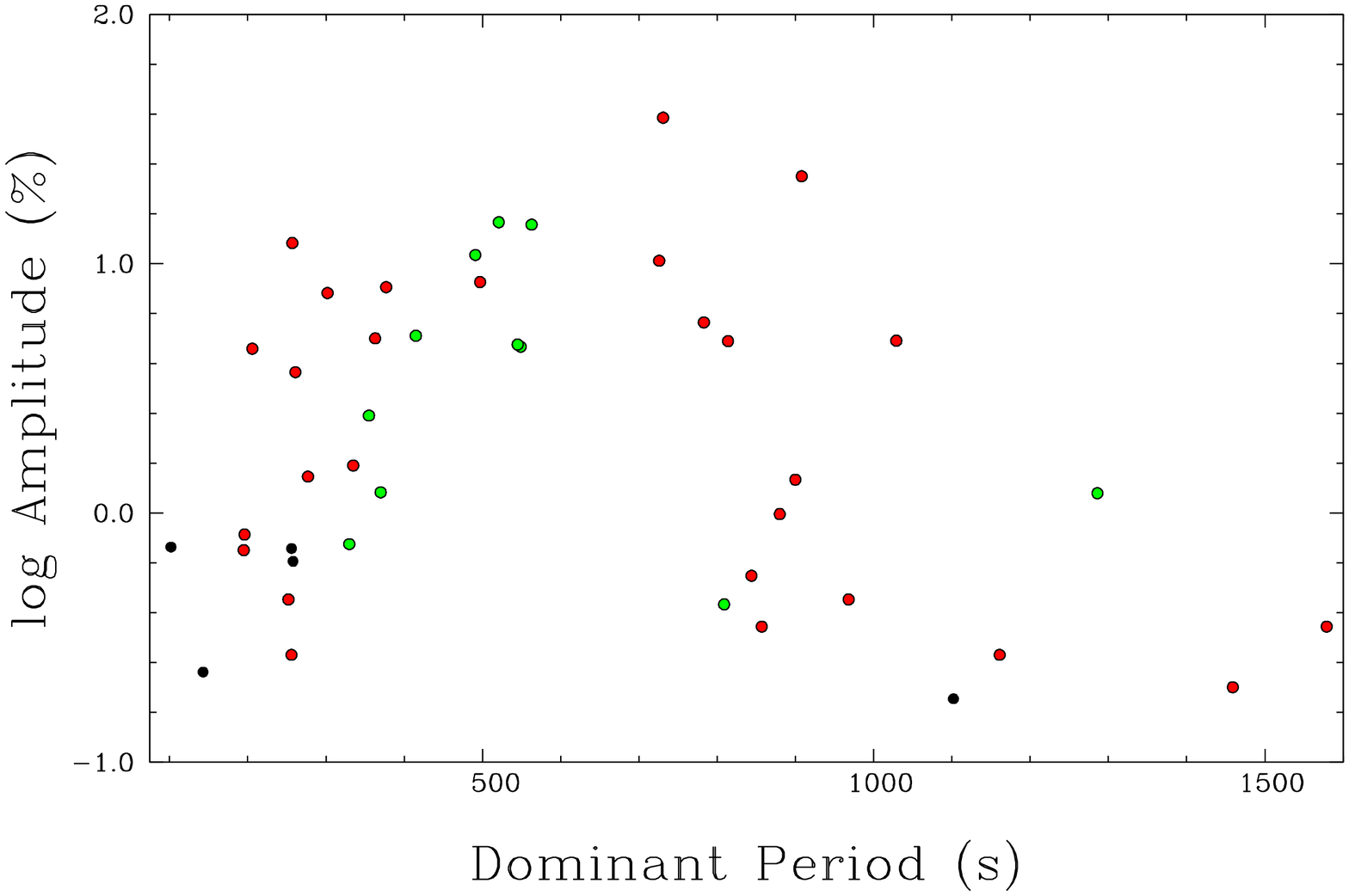}
\caption{Logarithm of the amplitude (in \%) against the dominant
  period for the new ZZ Ceti white dwarfs (red dots) and possible
  pulsators (black dots) in our sample; massive ($M>0.75$ \msun) ZZ
  Ceti stars are shown as green dots.}
\label{fig:pamp}
\end{center}
\end{figure}

Next, we look for a correlation between the amplitude and the dominant
period using the ZZ Ceti stars discovered in our sample. The results,
displayed in Figure \ref{fig:pamp}, reveal amplitudes varying from 0.2
to 35\%, wherein shorter periods show smaller amplitudes, followed by
an increase in amplitude until the dominant period reaches
$\sim$800~s, above which point the amplitudes start diminishing. Our
massive ZZ Ceti stars seem to follow the same overall trend as their
normal mass counterparts. Incidentally, this trend can be seen in the
$M-\Te$ plane of Figure \ref{fig:periods}, where amplitudes are at
their highest at the center of the strip, then diminish as the
pulsators move toward the edges. For our pulsators, higher amplitudes
also tend to coincide with light curves showing more complex
features. Overall, the ensemble characteristics observed here agree
with those established in the literature.  We did not, however, detect
any outburst events such as those described in \citet{hermes2017}.
This comes to no surprise as these events are known to last several hours,
while our observations were generally shorter than 2 hours.

We finish this section with a discussion regarding the authenticity of
our so-called possible pulsators. We compare their physical and
pulsational properties with those of the new ZZ Ceti stars in our
sample, starting with the warmest object. We have also included three
known ZZ Ceti stars in Figure \ref{fig:periods}, located extremely
close to the blue edge of the instability strip, to make up for the
lack of new pulsators within that region. These three known ZZ Ceti
possess some of the shortest periods and smallest amplitudes ever
detected: HS 1531$+$7436 with $P_d \sim111$~s and an amplitude of
$\sim$4~mma \citep{voss2006}, GD 133 with $P_d \sim120$~s and an
amplitude of $\sim$4~mma \citep{silvotti2006}, and G226-29 with $P_d
\sim100$~s and an amplitude of $\sim$1~mma \citep{kepler1983}. The
last two are relatively bright --- with {\it Gaia} magnitudes
$G=14.76$ and 12.29, respectively --- and their pulsations might not
have been detected if not for this. For instance, G226-29 had been
observed several times with telescopes as large as 1.6~m, but its
variability could not be confirmed until observations were secured
with the 6.5~m Multiple Mirror Telescope. We thus expect pulsators
very close to the blue edge of the strip to have periods around 100~s
and very small amplitudes, which is exactly the kind of weak signal we
detected in our possible massive pulsator J1207$+$6855. The rest of
the possible pulsators is located around 0.6 \msun\ in Figure
\ref{fig:periods}. Their periodograms mostly show peaks within
in the expected period range, although the amplitudes are too small
to be confirmed unambiguously. Their pulsations also follow the usual
period-amplitude trend, as shown in Figure \ref{fig:pamp}. All of our
possible pulsators will need to be re-observed with better
instruments, or at the very least, under exceptional observing
conditions. Space-based surveys (i.e., TESS and the upcoming 
PLATO 2.0 Mission; \citealt{plato2014}) may offer 
an interesting avenue to acquire higher-quality data, in particular 
for brighter objects. Furthermore, these surveys could also be useful
for asteroseismic studies of our new ZZ Ceti stars, as well as to verify 
with greater precision if NOV candidates are truly nonvariable.

\section{Conclusion}\label{sec:conc}

In this paper, we presented the first study of the photometric ZZ Ceti
instability strip using results from the combined {\it Gaia} and Pan-STARRS surveys. 
In addition to searching for new puslators, we aimed to verify whether ZZ
Ceti white dwarfs occupy a region of the $M- \Te$ plane where no
non-variable stars are found, supporting the idea that ZZ Ceti stars
represent a phase through which all hydrogen-atmosphere white dwarfs
must evolve.

We first selected all white dwarfs and white dwarf candidates in the
Northern Hemisphere within 100 parsecs of the Sun with parallax
measurements from the {\it Gaia} Data Release 2 catalog, which we then
cross-referenced with the Pan-STARRS Data Release 1. Using the
so-called photometric technique, we measured with high precision the
physical parameters ($\Te$ and $M$) of every object by combining
Pan-STARRS $grizy$ photometry with {\it Gaia} astrometry. Since the
Pan-STARRS photometry alone does not allow for a determination of the
chemical composition of white dwarfs, we also included SDSS or CFIS
$u$ photometry, when available, in our model atmosphere fits. The
$u$-band covers the Balmer jump, which represents a good discriminant
between hydrogen- and helium-rich atmosphere white dwarfs, and it can
be used efficiently to exclude non-DA stars from our list of ZZ Ceti
candidates. To establish a region of the $M- \Te$ plane where the DA
pulsators could be found, we first applied 3D corrections to the
spectroscopic parameters of a sample of bright ZZ Ceti stars. We also
made adjustments to the effective temperature of the boundaries of the
instability strip to account for the known discrepancies between
spectroscopic and photometric parameters, producing our final
empirical photometric instability strip. We identified a final sample
containing 173 ZZ Ceti candidates within this strip.

We acquired high-speed photometry for a sample of 90 ZZ Ceti
candidates within the photometric instability strip using the PESTO
instrument attached to the 1.6~m telescope at the Mont-M\'egantic
Observatory. Among these, 38 proved to be clearly variable, while 5
show possible small-amplitude pulsations, and 47 were not observed to
vary. Additionnally, 18 objects near, but above the blue edge of the
instability strip, were observed and showed no variability.

The implications of our findings, as well as the nuances of the
photometric technique in the context of ZZ Ceti identification, have
been discussed at length in this paper. The first remarkable result
was, of course, the large quantity of new ZZ Ceti white dwarfs
identified in our study. We discovered 11 massive ZZ Ceti stars
($M>0.75$ \msun), including two very rare ultra-massive pulsators,
making a significant contribution to the number of such known
objects. We attribute this high rate of identification of new massive
pulsators to the use of a volume-limited, rather than a
magnitude-limited, sample for the selection of our ZZ Ceti
candidates. The distribution of our new ZZ Ceti stars in the $M-\Te$
plane was shown to be in excellent agreement with our empirical
photometric instability strip, suggesting a good internal consistency
between the spectroscopic and photometric methods. The pulsation
ensemble characteristics of our sample in the $M-\Te$ plane were also
examined qualitatively, and showed good agreement with the empirical
trends previously established. In particular, massive pulsators seemed
to follow the same tendencies as their normal mass counterparts, with
the exception of the new ultra-massive variable white dwarfs.

We attempted to assess the purity of the instability strip by
investigating in depth the candidates showing no variability. Our
study turned out to be inadequate for a meaningful discussion of this
topic, and it will require further spectroscopic investigations of the
non-variable candidates. Observing the candidates located near both
boundaries of the strip with higher performance facilities than those
offered by the Mont-M\'egantic Observatory will also be necessary in
this context, as objects in these regions are known for their very
low-amplitude variations, and these are most likely not detectable
with our current means. Furthermore, high-speed photometric
observations of such objects will eventually allow us to constrain
more accurately the exact location of the boundaries of the
instability strip.

Finally, it would be interesting to apply this photometric approach to
identify new pulsating white dwarfs of different types. DBVs would
make an excellent choice, as they are the most studied class of white
dwarf pulsators besides the ZZ Ceti stars.

\acknowledgments

We thank N.~Giammichele for a careful reading of our manuscript and
for her constructive comments. We would also like to thank the staff
of the Observatoire du Mont-M\'egantic for their assistance and for
conducting queue mode observations. This work was supported in part by
the NSERC Canada and by the Fund FRQ-NT (Qu\'ebec). This work has made
use of data from the European Space Agency (ESA) mission {\it Gaia}
(https://www.cosmos.esa.int/gaia), processed by the {\it Gaia} Data
Processing and Analysis Consortium (DPAC,
https://www.cosmos.esa.int/web/gaia/dpac/consortium). Funding for the
DPAC has been provided by national institutions, in particular the
institutions participating in the {\it Gaia} Multilateral
Agreement. This research has made use of the NASA/IPAC Infrared
Science Archive, which is operated by the Jet Propulsion Laboratory,
California Institute of Technology, under contract with the National
Aeronautics and Space Administration. This work has also made use of
data obtained as part of the Canada-France Imaging Survey, a CFHT
large program of the National Research Council of Canada and the
French Centre National de la Recherche Scientifique. Based on
observations obtained with MegaPrime/MegaCam, a joint project of CFHT
and CEA Saclay, at the Canada-France-Hawaii Telescope (CFHT) which is
operated by the National Research Council (NRC) of Canada, the
Institut National des Science de l'Univers of the Centre National de
la Recherche Scientifique (CNRS) of France, and the University of
Hawaii.

\bibliography{ms}{}
\bibliographystyle{aasjournal}

\end{document}